# FROM THE BERLIN "ENTWURF" FIELD EQUATIONS TO THE EINSTEIN TENSOR I: October 1914 until Beginning of November 1915

Galina Weinstein

*Visitor Scholar, Center for Einstein Studies, Philosophy Department, Boston University*

*25 January, 2012*

I discuss Albert Einstein's 1914 review paper, "The Formal Foundation of the General Theory of Relativity" from two points of view: the main elements in the paper that appear to have led to the downfall of the Einstein-Grossman theory; and the elements that seem to have inspired Einstein during October 1915 to reformulate the 1914 Einstein-Grossmann theory in the form of the November 1915 and the 1916 General Theory of Relativity. First paper among three papers.

## I The Berlin "Entwurf" Theory

## 1 From Zurich to Berlin

### 1.1 August 1912-spring 1913: The Metric Tensor and the "Entwurf" Theory (in a nutshell)

Einstein wrote in his 1955 *Autobiographical Sketch*,[1]

"From the experience of the kind of scientific research which those happy Bern years have brought, I mention only one, the idea turned to be the most fruitful of my life. The special theory of relativity was several years old [...].

1909-1912, while I was teaching at the Zurich [1909-1910] and at the Prague universities of theoretical physics, I mused incessantly over the problem [gravitation]. 1912, when I was appointed to the Zurich Polytechnic, I was considerably closer to the solution of the problem. Of importance here proved to be Hermann Minkowski's analysis of the formal basis of the special theory of relativity".

Einstein found the appropriate mathematical tool used in Minkowski's formalism, the square of the line element:

$x^2 + y^2 + z^2 - (ct)^2$,

and in tensorial form, equation $(2)\ ds^2 = -\sum_\nu dx^2{}_\nu$. which is invariant under the Lorentz group.

Einstein then considered a non-flat four dimensional space-time, i.e., geometry of space-time which is curved. He recognized that the gravitational field should not be described by a variable speed of light as he had attempted to do in his 1912 Prague coordinate dependent theory of static gravitataional fields (his papers from February 1912, "Lichtgeschwindigkeit und Statik des Gravitationsfeldes", "The Speed of Light and the Statics of the Gravitational Fields", and March 1912, "Zur Theorie des statischen Gravitationsfeldes", "On the Theory of the Static Gravitataional Fields").[2]

Einstein realized that, "[…] the Gravitational field is described by a metric, a symmetric tensor-field of metric *gik*".[3] The metric tensor field is a mathematical object of ten independent components that characterizes the geometry of space and time.

Louis Kollros, a professor of geometry and mathematics at the ETH, who was a fellow student of Einstein, wrote in memory of Einstein in 1955,

"Even in Prague he had foreseen that his generalized theory of relativity demanded much more than the mathematics of the elegant special relativity. He now found that elegance should rather remain a matter for tailors and shoemakers.[4]

Einstein said that, "in 1912, I was looking for my old student friend Marcel Grossmann, who had meanwhile become a professor of mathematics in the Swiss Federal Polytechnic institute. He was immediately caught in the fire, even though he had as a real mathematician a somewhat skeptical attitude towards physics. When we were both students, and we used to exchange thoughts in the Café, he once said such a beautiful and characteristic remark that I cannot help but quoting it here: 'I admit that from studying physics I have benefitted nothing essential. When I sat on the chair earlier and I felt a little of the heat that came from my 'pre-seated' it grazed me a little. This has completely passed, because physics has taught me to consider writing that heat is something very impersonal'.

So he arrived and he was indeed happy to collaborate on the problem, but with the restriction that he would not be responsible for any statements and won't assume any interpretations of physical nature".[5]

Grossmann,[6]

"[…] looked through the literature, and soon discovered that the particular implied mathematical problem was already solved by Riemann, Ricci and Levi-Civita.[7] The entire development followed the Gaussian theory of curvature-surfaces, which was the first systematical use of generalized coordinates.[8] Reimann's achievement was the biggest. He showed how a field of *gik* tensors of the second differentiation rank can be formed."

[in his *Habilitationsschrift* on the foundations of geometry "Über die Hypothesen, welche der Geometrie zu Grunde liegen" ("On the hypotheses which lie at the foundation of geometry")].[9]

Pais recalls a discussion with Einstein in which he asked Einstein how the collaboration with Grossmann began."I have a vivid though not verbatim memory of Einstein's reply: he told Grossman of his problems and asked him to please go to the library and see if there existed an appropriate geometry to handle such questioning. The next day Grossman returned (Einstein told me) and said that there indeed was such a geometry, Riemannian geometry".[10]

Kollros added that, sometime upon his arrival, Einstein spoke about his concern with Grossmann and told him one day, "Grossmann, you have to help me, or I shall go crazy! And Marcel Grossman had managed to show him that the mathematical tools he needed, were created just in Zurich in 1869 by Christoffel in the treatise 'On the Transformation of the Homogeneous differential Forms of the Second Degree', published in Volume 70. of the 'Journal de Crelle' for pure and applied mathematics".[11]

Einstein wrote, "also I was fascinated by professor [Karl Friedrich] Geiser's lectures on infinitesimal Geometry, the true master pieces of art were pedagogical and helped very much afterwards in the struggle with the general theory of relativity".[12]

Einstein seemed to have had already a key idea of the Gaussian theory of curvature-surfaces from Geiser. In the Kyoto lecture Einstein is reported to have said,[13]

"This problem was unsolved until 1912, when I hit upon the idea that the surface theory of Karl Friedrich Gauss might be the key to this mystery. I found that Gauss' surface coordinates were very meaningful for understanding this problem. Until then I did not know that Bernhard Riemann had discussed the foundation of geometry deeply. I happened to remember the lecture on geometry in my student years [in Zurich] by Carl Friedrich Geiser who discussed the Gauss theory. I found that the foundations of geometry had deep physical meaning in this problem".

Einstein wrote about his switch of attitude towards mathematics in the oft-quoted letter to Sommerfeld on October 29, 1912,

"I am now occupied exclusively with the gravitational problem, and believe that I can overcome all difficulties with the help of a local mathematician friend. But one thing is certain, never before in my life have I troubled myself over anything so much, and that I have gained great respect for mathematics, whose more subtle parts I considered until now, in my ignorance, as pure luxury! Compared with this problem, the original theory of relativity is childish".[14]

And,[15]

"The problem of gravity was thus reduced to a purely mathematical one. Are there differential equations for *gik*, which are invariant under non-linear coordinate transformations? Such differential equations and *only* those were taken into consideration as the field equations of the gravitational field. The law of motion of material points was given by the equation of the geodesic line".

Einstein first struggled with these new mathematical tools in a small blue Notebook – named by scholars the "Zurich Notebook".[16] Einstein filled 43 pages of this notebook with calculations, while he was fascinated with Riemann's calculus.

While filling the notebook he received from time to time the new mathematical tools from Grossmann, and he wrote Grossmann's name in the notebook every time he got something new to indicate the tensors that he received from him. For instance, at the top of page 14L Einstein wrote on the left: "Grossmann's tensor four-manifold", and next to it on the right he wrote the fully covariant form of the Riemann tensor. It appears that Grossman suggested the four-rank Riemann tensor as a starting point.

On top of page 22R Einstein again wrote Grossmann's name.[17] He considered candidate field equations with a gravitational tensor that is constructed from the Ricci tensor; an equation Einstein would come back to in November 1915.

Einstein's collaboration with Marcel Grossmann led to two joint papers: the first of these was published before the end of June 1913, "Entwurf einer verallgemeinerten Relativitätstheorie und einer Theorie der Gravitation" ("Outline of a Generalized Theory of Relativity and of a Theory of Gravitation"[18], the "Entwurf" paper; and the second, almost a year later,[19] two months after Einstein's move to Berlin.[20]

The "Entwurf" theory was already very close to Einstein's general theory of relativity that he published in November 1915. The gravitational field is represented by a metric tensor, the mathematical apparatus of the theory is based on the work of Riemann, Christoffel, Ricci and Levi-Civita on differential covariants, and the action of gravity on other physical processes is represented by generally covariant equations (that is, in a form which remained unchanged under all coordinate transformations). However, there was a difference between the two theories. The "Entwurf" theory contained *different field equations* that represented the gravitational field, and these were *not* generally covariant.[21]

### 1.2 January 1913-May 1913: The decision to Bring Einstein to Berlin

Einstein was already collaborating with his school-mate Marcel Grossmann for a couple of months on the theory of gravitation. In January 1913 Fritz Haber wrote Hugo Krüss, a worker at the Prussian Ministry of Education, of his thoughts about bringing Einstein to the Keiser Wilhelm Institute in Berlin. Haber planned to bring Einstein to his institute of Physical Chemistry and Electrochemistry. Haber wrote to Krüss that "Einstein's closest colleague is Herr Planck, and then, Herr [Emile] Warburg and Herr [Heinrich] Rubens, who also work in disciplines related to his specialty".[22]

Einstein was still unaware of this state of affairs in Berlin. He was sitting at the ETH in Zurich working on his new gravitation theory. Einstein was troubled by the calculations he was just doing (on pages 19L-23L) in the Zurich Notebook.[23] He tried to extracted expressions of broad covariance from the Ricci tensor while imposing coordinate conditions. But then he was disappointed and he tried (on page 24R) to establish field equations while starting from the requirement of the conservation of momentum and energy. Still (on pages 25L and 25R) Einstein did not fully give up. He was trying to find a way to recover the new field equations from the "November tensor" (of page 22R) – the "presumed gravitational tensor" he wrote on this page.[24] But it was "Unmöglich" (impossible),[25] and in the next pages (26L and 26R) Einstein was led straight to the "Entwurf" field equations, which he also established through energy-momentum considerations; [26] the "Entwurf" paper was very likely ready for submission. The time now was late spring 1913.

At the same time, in Berlin, between January and May 1913, the emphasis seems to have shifted from Haber's original proposal. By late spring Max Planck and Walther Nernst had modified Haber's proposal, combining the idea of Einstein's membership in the Academy of Prussian Sciences with the prospect of his directorship of a Keiser Wilhelm Institute of Physics.[27] Eventually, Einstein would sit in Haber's institute until 1917, and a special membership in the Academy of Prussian Sciences was conferred to him.

**1.3 June 12, 1913: Einstein is formally proposed for Membership in the Prussian Academy**

To ensure Einstein's election to the Prussian Academy, the four most important German physicists and members of the Academy, Planck, Nernst, Rubens and Warburg submitted a request to the Prussian Ministry of Education.[28]

They first announced in the physical-mathematical class of the Prussian Academy of Sciences that they would submit a proposal for membership at the next session. The identity of the candidate was not given. Two weeks later they proposed Einstein for election as a regular member of the Academy.

Max Planck, the presiding secretary of the class, read aloud a text of proposal to the physical-mathematical class on June 12, 1913. The text said:[29] Einstein was born in March 1879 in Ulm, raised in Munich, is a citizen of Zurich from 1901, he was employed as a technical expert in the Patent office from 1902 to 1909. Only in 1905 he was awarded his doctoral degree at the University of Zurich, habilitated in 1908 in Bern, accepted an appointment as extraordinary professor of theoretical physics at the university of Zurich in 1909, and an ordinary professor at the German university in Prague the following year, and from there he came back to Zurich to the ETH in 1912.

Planck then said that by his papers in the field of theoretical physics, published for the most part in the *Annalen der Physik*, Einstein achieved at a young age, within the circle of his specialty, a worldwide reputation. His name became widely known in his famous treatise on the electrodynamics of moving bodies (1905), which established the principle of relativity.

However Planck told the attendants that fundamental as this idea of Einstein's has proved to be for the development of the principles of physics, its applications are for the present still at the very limit of the measurable. Planck mentioned Einstein's tackling of other central questions that proved to be much more meaningful for applied physics. He was the first to demonstrate the significance of the quantum hypothesis also for the energy of atomic and molecular motions, in that he derived from this hypothesis a formula for the specific heats of solid bodies.

Planck summarized that it can be said that among the major problems, with which modern physics is so rich, there is no one in which Einstein did not take a position in a remarkable manner. But added Planck, "That he might have in his speculations, occasionally, overshot the target, as for example in his light quantum hypothesis, should not be counted against him too much; because without taking a risk, even in the most exact science, one is not driven to real innovation".

Planck added with regard to the theory of gravitation, "At the moment he works intensively on a new theory of gravitation; with what success, only the future will tell".

Planck ended his proposal by saying that he and the three undersigned members (Nernst, Rubens, and Warburg) are aware that their proposal to accept so young a scholar as a full member of the Academy is unusual; but they believe not only that the unusual circumstances adequately justify the proposal, but also that the interests of the Academy really require that the opportunity that now manifests itself to obtain such an extraordinary power be taken in its full possibility.

Planck thought that even if they could not guarantee the future, they could be convinced that the recommended previous accomplishments of the nominee fully justify his appointment to the most distinguished state scientific institute; and there were also further evidence about the fact that the entire world of physics would consider Einstein's entering the Berlin Academy of Sciences as an especially gain for the Academy

Einstein also worried about the future. Kollros recalled that before Einstein left to Berlin "I accompanied him home that night. He said: 'The Berliner gentlemen speculate that I am an award-winning Chicken-hen, but I do not know if I can still lay eggs!' "[30] Einstein also felt he was collected like rare stamps.[31]

Planck had the right intuition. Already in 1905 he accepted Einstein's relativity paper for publication in the *Annalen*, and he was the first known scientist to respond to it in 1906. Planck had the feeling that the principle of relativity was the right direction. Now Planck thought that Einstein was the right person. He would not always believe in Einstein's new theory of gravitation, but his first intuition was to bring him to Berlin.

Walter Nernst took the occasion of the class meeting of the 12 June to make a confidential announcement that Einstein's salary might be raised to 12,000 marks, because of a personal commitment by Leopold Koppel (a financer from the board of directors of Haber's institute) to provide 6000 marks a year for twelve-year period to the Academy as donation; the latter expressed his willingness to Haber to help defray the costs of an appointment for Einstein in January 1913.[32] Planck said in this meeting that even Carl Schwarzschild explained the application to two (probably esteemed members of the Academy) Herren Fischer and Waldeyer, to contest them, and told them that it is more favorable and worthy if the Academy would pay the full 12,000 marks, while Herr. Koppel was obliged to pay his part.[33] Koppel would pay four quarterly installments of 1500 marks each into a fund of the physical-mathematical class at the finance office of the University of Berlin.[34] Schwarzschild would later be the first to give an exact solution to Einstein's generally covariant field equations.

Meanwhile in Zurich Einstein was fully immersed in his science, his newly "Entwurf" gravitation theory. His joint paper with Grossman – the "Entwurf" paper – was published before the end of June 1913.[35] In June 1913 Michele Besso visited Einstein in Zurich, and they both tried to solve the new "Entwurf" field equations to find the perihelion advance of Mercury in the field of the static sun in the Einstein-Besso manuscript.[36] At the same month, Einstein had another visitor in Zurich. Paul and Tatyana Ehrenfest came from Leyden to a trip to Zurich. They stayed in a pension in Zurich, but they spent a great deal of time with Einstein and his family. Ehrenfest met Marcel Grossmann during this stay in Zurich, and he also met Besso.[37] Einstein was thus surrounded by two sounding boards – Besso and Ehrenfest – while he tried to find solutions to the Einstein-Grossmann field equations and solve the problem of the precession of the perihelion of Mercury.

On July 3, 1913, the physical-mathematical class had voted twenty-one to one in Einstein's favor, and a week later the plenary session debated the issue.[38]

### 1.4 July 11-15, 1913: Planck's and Nernst's visit to Zurich

Planck and Nernst left Berlin for Zurich on the night train of Friday, 11 July, 1913, at the suggestion of Koppel. They went to Zurich personally to influence Einstein in favor of their plan.[39] They came to Einstein who was younger than them – 34 years

old – but already universities all over the world offered him positions; even though Einstein then was only eight years in the academic world.[40]

Planck had asked Einstein what he was working on, and Einstein described general relativity as it was then (the "Entwurf" theory). Planck said: "As an older friend I must advise you against it for in the first place you will not succeed; and even if you succeed, no one will believe you".[41]

Planck and Nernst's came to Einstein with the following plan:[42]

A research institute for physics that did not yet exist, and there was no chance that such institute would be founded in the near future. But Einstein would be (in 1917) the first director of this institute in preparation, while in the meantime he would lead the research in other institutes, where physics has been carried in areas under his instruction.

Einstein was also to become a member of the Royal Prussian Academy of Science in Berlin. It was considered a great honor to be a member of this corporate body, and many significant professors of the University of Berlin all their lives never achieved it. The membership was not a profession for most (of the incumbents), they could not live on it, but an honorary position. There were a few positions that were fitted by foundations (Koppel's) with such a large salary, so that they could be exercised full-time. Such a position was offered to Einstein. And moreover, Einstein was the youngest member of this prestigious academy.

Both in the Academy and in the Keiser Wilhelm Institute his main professional occupation should be the organization of research. He was also to have the title of professor at the University of Berlin, but there were no obligations, except the right of lecturing as much or as little as he liked. He should have nothing to do with the administrative work of the university, with examination, with the appointment of new professors, etc. His only obligation was to be active in organizing research in the academy and the Keiser Wilhelm Institute. He could teach and engage in research as much as he liked.

Einstein would earn a regular salary of 900 deutsche Marks per year, and in addition to this the Prussian Academy of Sciences approved later on November 22, 1913 the special personal salary of 12,000 deutsche Marks per year; both salaries would be paid to him beginning with the first of the month in which he moved to Berlin.[43] There were great advantages offered by this invitation. Besides the academic honors that the Prussian Academy bestowed on Einstein, he was about to move to Berlin with a much larger salary than he had in Zurich.

Louis Kollros recollects memories about the visit of Planck and Nernst to Einstein: "Max Planck and Walter Nernst made great efforts to bring Einstein to Berlin. They came to Zurich, to make him offerings; he could come to do physical research at the Kaiser-Wilhelm-Institute organization, which was created in 1911 by Kaiser Wilhelm II. He would become a member of the Academy of Sciences, could teach at the university if he wished, and have all leisure to use the time for his own research. The democrat Einstein loved Switzerland, he at first hesitated to reply to the Berlin coveted agreement.[44]

During their visit to Zurich Planck and Nernst made a Sunday excursion on July 13, 1913, and Einstein thought their offer over.

Frank says that despite Einstein's unusual talents he could still expect to be stimulated by new ideas, since it was always fruitful to receive the criticism of so many scientists capable of independent thinking working in many different fields.[45] He would thus have as much of an opportunity as he desired to come into contact with the many leading physicists, chemists, and mathematicians who were in Berlin. This was a great advantage, because Berlin was the center of theoretical physics at that time.

On the other hand, it was difficult for Einstein to decide to return to the center of that Germany from which he had fled as a pupil. He had renounced his German citizenship two decades earlier, preferring being stateless and then Swiss to the military service in Germany. It seemed to him even a kind of betrayal of his convictions to become a member of a group with which he did not harmonize in so many respects, simply because it was connected with a pleasant position for himself. It was for him a struggle between his personality as a scientific investigator who could benefit by moving to Berlin, and his feeling as a member of a certain social group.[46]

In addition there were also personal factors that entered into the decision. Einstein had an uncle Rudolf Einstein in Berlin a fairly successful businessman. In 1898 when Einstein was young his father Hermann founded an electrical factory in Milan. Rudolf Einstein was persuaded to finance the enterprise, even though he had lost money in such earlier ventures. For a time around 1900 Hermann Einstein seems to have been free of financial worries. Rudolf had a daughter Elsa, who was now a widow. Einstein remembered that his cousin Elsa, as a young girl, had often been in Munich and had impressed him as a friendly, happy person, and he remembered his uncle's generousity. The prospects of being able to enjoy the pleasant company of this cousin in Berlin enabled him to think of the Prussian capital somewhat more favorably.

Einstein later wrote Elsa, on December 2, 1913, "You can hardly imagine how much I look forward to this spring, first of all because of you, but also because of Haber and Planck. The latter has shown to me really touching friendship, and you know the same from the first. But what is most important for me is that I have someone with whom I can talk on human things, and with whom I can be personal, and this is you".[47] Hence, it was first the company of Elsa, and then that of Planck and Haber, who drew Einstein to Berlin.

And so Einstein decided to accept the offer. Einstein met Planck and Nernst at the train station in Zurich and indicated his visitors by a prearranged signal – the waving of a white cloth; he accepted the offer.

He wrote Elsa on Monday, July 14 about the visit and the offer, and told her that "next spring at the latest, I'll come to Berlin forever. It is a colossal honor that has been there bestowed upon me, because I will be Vant Hoff's successor".[48] Jacobus Vant Hoff has been granted the first research professorship in the Prussian Academy that included an annual salary of 10,000 marks and an honorary professorship at the University of Berlin, free of teaching obligations. He died in 1911, and his chair had been vacant since then.[49] Indeed Einstein was to receive 12,000 marks as compared to Vant Hoff's 10,000, and like him, to have no teaching obligations.[50]

Einstein told Elsa thereafter, "Don't tell anyone anything about the matter. It still requires a decision of the Plenum of the Academy, and it would look bad if something like that were to become public knowledge".[51] The Plenum of the Academy still did not elect Einstein, and at this stage he only gave Planck and Nernst a positive reply. So Planck and Nernst left Zurich for Berlin on the night train of Monday and arrived home on the morning of Tuesday, the 15th.[52]

### 1.5 July-August 1913: Einstein is elected to Academy and Elsa consults with Haber on Einstein's appointment

Einstein waited a few days, and probably did not get any reply from Elsa. He then wrote her again on Saturday, July 19, 1913, "In case you have not received my first letter [from Monday], I share with you again that I have been elected to the Prussian Academy in a very similar way as Vant Hoff, with a sufficient salary and without any further obligation, so that I can devote myself to scientific work".[53] Einstein also reported to his friend Jakob Laub on Tuesday, July 22, 1913 that on Easter he will be going to Berlin as an Academy-man without any obligations.[54]

Before Thursday, Einstein wrote Elsa again, and told her that "it is not yet certain".[55] It was not certain, because the final vote on Einstein's membership in the Prussian Academy did not yet take place.[56] But finally on Thursday, July 24, 1913, the plenary session of the Prussian Academy of Sciences elected Einstein as a member by the final vote of forty-four to two. The result was reported to the Prussian Ministry of Education four days later.[57] It was already almost August 1913.

Shortly after Planck and Nernst's left Zurich, Besso visited Einstein in Zurich again in August 1913, and they continued to solve the "Entwurf" field equations to find the advance of the Perihelion of Mercury. At about the time when Besso visited Einstein – Einstein was dissatisfied with his "Entwurf" field equations.[58]

Einstein wrote Elsa on August 11, 1913, "your assistance with my appointment has probably not been so ineffective: Haber knows his Pappenheimer. He knows already how to estimate the influence of a friendly cousin".[59]

### 1.6 November-December 1913: resignation from ETH and Acceptance of election to Academy

Einstein was busily working on his theory of gravitation. After August 1913 he spoke in two conferences during September 1913: He gave a lecture before the annual meeting of the Naturforschende Gesellschaft in Frauenfeld on September 9, 1913, and at the 85th congress of the German Natural Scientists and Physicists in Vienna (September 23, 1913).[60] Both talks were dedicated to his new gravitation theory. He prepared the latter talk during August, and it was too long.[61] His colleagues – especially from the Göttingen school – attacked his new theory because his gravitational equations were not generally covariant. But on November 2, 1913 he reported to Hopf that "it can easily be proved that a theory with generally covariant [field] equations cannot exist".[62] Einstein already possessed the Hole Argument; and he also had the new offer in Berlin.

On October 7, 1913 Planck sent a letter to a possible influencing professor who requested material in order to scientifically assess Einstein's candidacy for the position in Berlin. The recipient of the letter could not be located. Planck sent him the material and the following supporting letter on behalf of Einstein. Planck took exactly the first sections from the supporting letter that he had read earlier aloud to the physical-mathematical class on June 12, 1913,[63] and reproduced them in the letter to the professor. The June 1913 letter was also signed by Nernst, Rubens and Warburg; but in light of the similarity to Planck's October, 1913 letter to the unknown professor, it is possible that Planck had also written the June 1913 letter read to the physical-mathematical class.

Here are the main sections of Planck's letter:[64]

Sehr geehrter Herr Professor!

Auf Ihren Wunsch stelle ich Ihnen gern noch einiges Material zur Verfügung, das für die wissenschaftliche Würdigung des Herrn Prof. Einstein Benutzt werden könnte.

Schon mit jungendlichen Jahren hat sich Einstein durch seine Arbeiten auf dem Gebiet der theoretischen Physik einem Weltruf erworben. Am meisten ist sein Name bekannt geworden durch das von ihm 1905 in seiner Abhandlung über die Elektrodynamik bewegter Körper aufgestellte Prinzip der Relativität, […]

So Fundammental sich dieser Gedanke Einsteins für die Entwicklung der Physikalischen prinzipien erwiesen hat, so liegen doch die Folgerungen desselben einstweilen noch hart an der Grenze des Meßbaren".

"Einstein achieved at a young age, by his papers in the field of theoretical physics, published for the most part in 1905, a worldwide reputation, His name became widely known in his famous treatise on the electrodynamics of moving bodies (1905), which established the principle of relativity […]

So fundamental as this idea of Einstein's has proved to be for the development of the principles of physics, its applications are for the present still at the very limit of the measurable".

On November 30, 1913 – Einstein sent his resignation letter to Robert Gnehm, the president of the Federal School Council of the ETH. Einstein told him that he got an offer from the Prussian Academy of Sciences, and he requested to have him released, effective April 1, 1914.[65] On December 6, 1913 a meeting of the Swiss Federal Council gathered in order to approve Einstein's request. Robert Gnehm reported to the meeting that he tried to persuade Einstein not to resign his position, but he understood that the position in Berlin was important to Einstein, given the fact that it was free of any teaching obligation. Two days after the meeting, on December 8, 1913, Gnehm addressed a request for approval of Einstein's resignation after detailing the inducements offered to Einstein to stay – an increase in salary, lifetime tenure, elimination of any teaching obligation, and occasional grants from an ETH research fund.[66]

But a day before that – on December 7, 1913 – Einstein had already wrote the Prussian Academy of Sciences that he declares "herewith that I accept this election".[67] On December 10, 1913, the Prussian Academy notified the Ministry of Education of Einstein's acceptance of the election, and the plenary session of the Prussian Academy was notified on December 18, 1913. Einstein informed the Academy that he would like to take up his new duties during the first days of April 1914.[68]

On December 11, 1913 the Swiss Federal Council approved, upon Einstein's request, his release from the professorship at the Polytechnic of Zurich. Robert Gnehm informed Einstein on December 15, 1913 about this approval.[69]

Louis Kollros recalls that as Einstein received the final appointment, he decided to leave Zurich and his family [Mileva and his two sons eventually returned to Zurich]. [Soon after the Vienna talk] On 13 November 1913 he sent the president of the board of education of the ETH his resignation letter [which took effect on April 1, 1914]. He quitted the Winter semester 1913/14 [which ended on March 21, 1914[70]]. At the end of December we organized a farewell dinner in "Crown Hall" [Kronenhalle] in Zurich. We all felt sorry he was leaving. He was delighted to be able to devote all his time to his research, delighted but somewhat anxious, because he did not know what the future will bring him.[71]

### 1.7 January-March 1914: Difficulties with Keiser Wilhelm institute for Physics in Berlin

Until January 1914 there was still no Keiser Wilhelm institute for Physics in Berlin. But at the beginning of January a new initiative was developed during a meeting held at the Prussian Ministry of Education. Max Planck suggested that one should wait with the creation of an institute "until he [Einstein has] gotten used to conditions in Berlin". In the beginning of January 1914, Walter Nernst proposed as one alternative the creation of a scientific committee to supervise research, composed of Planck, Einstein, Haber, Warburg, Rubens, Max von Laue and Ernst Beckmann.[72]

Einstein hoped that the question of a physics institute would be resolved upon his arrival in Berlin. However, he told Elsa a month later in February that for the time being nothing has become of the institute. [73]

Already a year before, in late January 1913, Friedrich Schmidt of the Prussian Ministry of Education believed that a theoretical physical institute was "unnecessary"; i.e., it was not necessary to establish an institute for Einstein, because he is not an experimentalist; this could explain the postponing of the establishing of the institute for Einstein.[74]

In early February 1914 Nernst, Haber, Planck, Rubens, and Warburg circulated a proposal to the Prussian government, to the Kaiser Wilhelm Society, and to the Koppel Foundation for the creation of the Institute of Physics Research of the Keiser Wilhelm Society. Based on Nernst proposal of early January 1914, and on the need to hold down overhead costs, it suggested that the institute occupy modest quarters, that experimental work would be conducted in already existing laboratories, and that the administrative burden would be placed on a scientific committee, of which Einstein was proposed as "permanent honorary secretary".[75]

In late March 1914 the Keiser Wilhelm society discussed an offer by the Koppel foundation to support the opening costs of the institute, and architects were consulted on its construction. Einstein told his friend Zangger on the same month, "There I will probably get my own institute and an assistant".[76]

Einstein was already worried about this when he wrote Lorentz in August 14, 1913, "I do not know right now what to do, because I have neither an institute nor an assistant in Berlin".[77] And Einstein also wrote Ehrenfest later in November 1913 that with regards to his appointment in Berlin, he was not allowed to take an assistant. But since he loves to work in collaboration with others he would request an assistant. He thought of Adrain Fokker, Lorentz's student.[78] Fokker was working at that time with Einstein in Zurich (during the Winter semester of 1913-1914). And with Fokker he would discover the covariant formalism of Nordström's theory early in 1914.[79]

However, Einstein's membership in the Prussian Academy did not confer the right of having an assistant. Einstein first needed to receive the directorship of an institute, and only then he could get an assistant.[80]

So without an institute and without the ability to take an assistant, Einstein would rather be left with his best and eternal friends, Michele Besso, Paul Ehrenfest, and so on, and of course with his loyal colleagues, Lorentz, Freundlich, Sommerfeld, and others.

**1.8 March-April 1914: Departure from Zurich to Berlin**

Einstein was 35 years old on March 14, 1914, and he left Zurich shortly afterwards on March 20 or 21, 1914; at first he told Elsa that he would arrive around the 10$^{th}$ of April, but before that he planned going on the 8$^{th}$ of March to a congress in Paris, which he finally did not go to. He thus arrived at Berlin earlier on April 6, 1914.[81]

Before coming to Berlin he went to Holland,[82] where he spent a week between 23 and 29 of March in Leyden and Haarlem. He went to visit his best friends, Lorentz and Ehrenfest. He wanted to talk with them on gravitation. The work with gravitation progressed, but it was under extraordinary efforts, because gravitation became fragile. Einstein was still troubled by the restricted covariance of the field equations. He told Ehrenfest that he was also writing Lorentz.[83] Einstein made a trip to Leyden with Lorentz's former student Fokker.[84]

Einstein's wife Mileva went with his children around New Year to Fritz and Clara Haber while searching for an apartment in Berlin.[85] Mileva had found an apartment on the trip to Berlin very close to Haber's institute in Berlin-Dahlem, and thus not so far from Freundlich; the apartment was located at Ehrenbergstraße 33 in Berlin-Dahlem.[86] She later departed Zurich at the end of March but returned in mid-April.[87] Einstein arrived alone to Berlin at the beginning of April.

This marked the end of the effective collaboration between Einstein and Marcel Grossmann. Einstein wrote towards the end of the *Skizze*,[88]

"While I was busy at work together with my old friend, none of us thought of that tricky suffering, now this noble man is deceased. The courage to write this little colorful Autobiographical Sketch, gave me the desire to express at least once in life my gratitude to Marcel Grossmann".

In Berlin Einstein sat in the Prussian Academy of Sciences as a young man among older men, usually proud, authoritative, sometimes having great achievements.[89]

Einstein told Besso that he was going to live in Dahlem and have a room (office) at Haber's institute, the Kaiser Wilhelm Institute of Physical Chemistry and Electrochemistry.[90] "So for me", said Einstein, "it would be the beginning of (God help me [Gottseibeimir]) a Berlinerization [Verberlinerung]". [91]

Einstein gave an inaugural address to the Leibniz-session of the Academy on July 2, 1914, "Antrittsrede in der Prussischen Akademie der Wissenchaften". The talk was general and dealt with the principles of theoretical physics. Towards the end of his talk he mentioned briefly, "It turns out that you arrive at a definite extension of the relativity theory, if one essentially specifies a principle of relativity in this extended sense [for non-uniform motions]. One is quickly led thereby to a general theory of gravitation, which includes dynamics".[92]

Planck gave a reply speech to this inaugural speech. On July 7, 1914 Einstein wrote Planck a private letter that was a reply to a remark that Planck had expressed in this speech.[93] Planck was already skeptic before about Einstein's quantum hypothesis; and now he was also skeptic about Einstein's new gravitation theory. Planck heard Einstein speaking about extending the principle of relativity, thereby arriving quickly at a definite theory of gravitation. He felt that Einstein was willing too quickly to renounce his most basic principle from his special theory of relativity, namely, the principle of the constancy of the velocity of light. Planck who spoke in June and October 1913 about Einstein's special theory of relativity with unflagging enthusiasm, could not accept Einstein's renouncing of the principle of the constancy of the velocity of light.

Einstein tried to explain to Planck that he did not actually renounce the principle of the constancy of the velocity of light. He told him that the two heterogeneous conditions on $g_{\mu\nu}$ were:

1) The analogue to Poisson's equation.

2) The requirements which enable the introduction of a constant c.

Thus, Einstein requested correspondence with Newton's theory and with the special theory of relativity as a limiting case.

### 1.9 Spring-September 1914: the solar eclipse expedition that never tested the "Entwurf" theory

Einstein wrote Zangger, "Objectively, I must say that I am doing the right thing by going there, I believe that the personal relationships could prove fruitful to many colleagues there. Especially the astronomers are important to me (at present)".[94]

Erwin Freundlich – who had a post in the Berlin observatory – was in Berlin waiting to demonstrate Einstein's prediction regarding the bending of light rays in the gravitational field of the sun. And Einstein very likely thought that his "personal relationships could prove fruitful" to Freundlich, and would assist in realizing the expedition mission if he was in Berlin.

In spring 1914 it was not only Elsa and Planck who drew Einstein to Berlin. It was also Freundlich in Berlin, and Einstein wanted "to follow the thing from nearby".[95] He wished to take care that the expedition would be performed and his new theory of gravitation be approved by experiment.

The outbreak of World War I in August 1914 forced complications for the solar-eclipse expedition and also for the Keiser Wilhelm Physics institute:

The postponement of the decision, but the institute was finally set up in 1917,[96] long after Einstein had set up his generally covariant field equations; and the verification of the deflection of light in the gravitational field had quite the same fate.

On August 21, 1914 there was a total eclipse of the sun whose path totality passed through Feodesiya in southern Russia. On the first of August Germany declared war on Russia and World War I had broken out a day afterwards.

Erwin Freundlich had planned to mount an expedition to test Einstein's prediction of the deflection of light in the gravitational field of the sun already in 1913, and he had discussed the plans in correspondence with Einstein.[97] The Berlin observatory and its director Hermann Struve refused to provide the funds. Einstein thus had to provide them from private sources. Eventually the funds came from the chemists Emil Fisher and from the Krupp Foundation.[98]

On July 19, 1914, Freundlich left Berlin with his colleague Walther Zurheilen accompanied by a technician and other members of the Royal Prussian Observatory. After a week of travel they arrived at Feodosiya in the Crimea on July 25, where they met the well-equipped expedition of the Argentine observatory of Córdoba. Freundlich had borrowed parts of telescope from the Argentine observatory and had assembled four cameras altogether. Freundlich and his German team were converted from visiting astronomers to enemy aliens.[99]

On August 4, Freundlich was interned as a prisoner of war in Odessa and his instruments impounded. Einstein in Berlin was worried and wrote his friend Ehrenfest, "My good astronomer Freundlich will experience captivity instead of the solar eclipse in Russia. I am worried about him".[100] However, a few weeks later on August 29 Freundlich was exchanged along with other Germans for Russians officers in similar circumstances, but was forced to leave his instruments behind (equipment that consisted of a telescope valued at 20,000 marks, a tent, 975 marks, 2 chronometers, 700 marks, meteorological instruments, 261 marks, and miscellaneous items, 150 marks).[101]

Freundlich returned to Berlin by the end of September 1914, just in time to read Einstein's new October 1914 "Entwurf" review article.[102] The hectic times from October 1914 until November 1915 is discussed now.

## 1.10 Einstein First Review Article: "The Formal Foundation of the General Theory of Relativity" – October 1914

Einstein's first big project on Gravitation in Berlin was to complete by October 1914 a summarizing long review article on his "Entwurf" theory. The paper was published in November of this same year. This version of the theory was an organized and extended version of his "Arbeits" (works) with Grossmann, the most fully and comprehensive "Entwurf" theory of gravitation; a masterpiece of what would finally be discovered as faulty field equations. The paper is written in somewhat archaic or not always conventional mathematical language; by this it reveals Einstein's initial trials with the new mathematical tools that he had received from Grossmann, and his manipulating these tools while sometimes giving them new names and his struggles with these tools.

For instace, Einstein had to use the special relativistic limit and weak fields approximation, and assume that space was flat in order to obtain the Newtonian results. Einstein reasoned that in the general case, the gravitational field is characterized by ten space-time functions of the metric tensor. $g_{\mu\nu}$ are functions of the coordinates ($x_\nu$). In the case of special relativity this reduces to $g_{44}=c^2$, where c denotes a constant. Einstein took for granted that the same degeneration occurs in the static gravitational field, except that in the latter case, this reduces to a single potential $g_{44}=c^2$, where $g_{44}=c^2$ is a function of spatial coordinates, $x_1$, $x_2$, $x_3$.[103]

In a letter to Arnold Sommerfeld dated July 15, 1915, Einstein wrote, "Grossman will never claim to be considered a co-discoverer. He only helped in guiding me through the mathematical literature, but contributed nothing of substance to the results."[104] Einstein thus explained that the "Entwurf" theory was his theory; and two months later Einstein took full responsibility, and eventually he blamed only himself, when the "Entwurf" field equations collapsed.

### 1.10.1 A. "The Basic Idea of the Theory"

**In section §1**, "Introductory Considerations", Einstein considered two systems: one K, which is in uniform translation motion, and is a legitimate Galilei-Newtonian coordinate system, and the other K', which is in uniform rotation relative to K. Centrifugal forces then act on the masses at rest relative to K', while they do not act upon the masses which are at rest relative to K.

Newton already saw in this a proof that the rotation of K' had to be regarded as "absolute", and thus one could not consider K' as "at rest" like K. This argument, however, wrote Einstein – as shown particularly by Ernst Mach – is not valid. The existence of these centrifugal forces do not necessarily require the motion of K'. We could just as well derive them from the averaged rotational movement of distant masses in the environment with respect to K', and thereby treating K' as "at rest".[105]

Einstein said that, if Newtonian laws of mechanics and gravitation did not allow for such a view, it could very well be rooted in defects in this theory. At that time Einstein was under the impression that the above argument spoke in favor of his new theory of general relativity: the centrifugal force, which acts under given conditions upon a body, is determined by precisely the same natural constant as the effect of the gravitational field, such that we have no means to distinguish a "centrifugal field" from a gravitational field. Einstein thus interpreted the rotating system K' as *at rest* and the centrifugal field as a gravitational field.

Einstein introduced the inertio-gravitational field. Einstein said that it "reminds the view of the original (special) theory of relativity [ursprünglichen spezielleren Relativitätstheorie] in that the motion of an electric mass in a magnetic field, acted upon by a pondermotive force, can also be regarded as the effect of that electric field, which exists from the standpoint of the moving reference system of the mass, at the location of the mass". [106]

Einstein was referring to his magnet and conductor thought experiment. He considered the existence of the electric field to be relative and dependent on the state of motion of the reference system. Only the electric and magnetic field combined together could have an objective reality. This phenomenon – as he said later in a 1921 unpublished draft of a paper for *Nature* magazine, "Fundamental Ideas and Methods of the Theory of Relativity, Presented in Their Development" – forced him to postulate the special principle of relativity.[107]

Einstein realized that, "The gravitational field is considered in the same way and has only relative existence, like the electric field generated by magneto-electric induction".[108] The difference between the gravitational field and the electromagnetic field is: in the first case there is breakup into inertia and gravitation relative to the acceleration, and in the latter case the breakup is into electric and magnetic fields relative to the velocity. One cannot have unification of inertia and gravitation because they depend on acceleration. [109]

Einstein very likely had this realization already in Bern when he thought about the imaginary man falling from the roof and the equivalence principle, [110]

"The gravitational field is considered in the same way and has only a relative existence like the electric field generated by magneto-electric induction. *Because for an observer freely falling from the roof of a house there is during the fall* – at least in his immediate vicinity – *no gravitational field*. Namely, if the observer lets go of any bodies, so they remain relative to him, in a state of rest or uniform motion, regardless of their particular chemical and physical nature. The observer is therefore justified in interpreting his state as being 'at rest'".

In 1914 he thought about the equivalence principle in terms of the inertio-gravitational field. This realization was already a mature equivalence principle.

**In section §2**, "The Gravitational Field", Einstein wrote the equations of motion of a material point in a gravitational field. A material point in a gravitational field moves according to the formula:

$$(1)\ \delta \int \{ds\} = 0,$$

and,

$$(2)\ \mathrm{ds}^2 = -\sum_\nu \mathrm{dx}^2{}_\nu.$$

where, $x_1 = x$, $x_2 = y$, $x_3 = z$, $x_4 = it$. Einstein called ds the local time "eigenzeit". The variation in (1) is formed so that the coordinates $x_\nu$ at the end points of integration remain unchanged. [111]

Einstein then replaced the line element of Minkowski's flat space-time (2) by the more general form of the line element:

$$(2a)\ ds^2 = \sum_{\mu\nu} g_{\mu\nu} dx_\mu dx_\nu.$$

The ten quantities $g_{\mu\nu}$ determine the gravitational field and they are functions of $x_\nu$.[112] Einstein said that equation (1) remains valid under any coordinate transformation. The movements of a point mass in a gravitational field are characterized by the variational equation (1) and are represented by geodesics of the space-time under consideration.

The special theory of relativity is still valid in the infinitesimally small local region,

$$(2b)\ ds^2 = \sum_{\mu\nu} g_{\mu\nu} dx_\mu dx_\nu = -\sum dX_\nu^2\ .$$

$X_\nu$ ($\nu$ = 1, 2, 3,4) are coordinates of space and time in the infinitely small four dimensional local regions. They are measured with unit-measuring rods and a suitably chosen unit-clock. Einstein called the quantity $ds^2$ the "naturally measured distance" (natürlich gemessene Abstand) between two space-time points. [113] He then wrote, "In the general theory of relativity ds plays the same role as the element of a worldline in the original theory of relativity".[114]

### 1.10.2 B. "From the Theory of Covariants"

**In sections §3 to §6** Einstein presented in a systematic and ordered manner the basic methods of tensor calculus, absolute differential calculus that he had used in his "Entwurf" papers until then. As can be seen from these sections Einstein's collaboration with Grossman assisted him very much from the mathematical point of view. In sections §3 to §6 Einstein filled the gap of Grossmann's "mathematical part" that appeared in his two previous papers with Grossmann, and instead of Grossmann he gave a preliminary mathematical part to his theory. By doing this, in the 1914 review paper Einstein set a new important standard for future textbooks on the general theory of relativity. In most textbooks on the general theory of relativity the same format of presentation appears: first, the basic methods of tensor calculus are presented, after which there is a systematic teaching of general relativity.

**In section §4**, "Contravariant Tensors", Einstein wrote the transformation law for contravariant tensors:

$$(8)\ A^{\mu\nu\prime} = \sum_{\alpha\beta} \frac{\partial x'_{\mu}}{\partial x_{\alpha}} \frac{\partial x'_{\nu}}{\partial x_{\beta}} A^{\alpha\beta}$$

Einstein was going to refer to this equation in the paper and afterwards in his correspondence with his colleagues.[115]

[and for covariant tensors:[116] $(5a)\ A'_{\mu\nu} = \sum_{\alpha\beta} \frac{\partial x_{\alpha}}{\partial x'_{\mu}} \frac{\partial x_{\beta}}{\partial x'_{\nu}} A_{\alpha\beta}$].

Following Ricci and Levi-Civita, Einstein explained in section §3 that he represented contravariant components of a tensor by raised indices and covariant components by lowered indices.[117] In section §6 Einstein wrote some relations concerning the (metric) fundamental tensor $g_{\mu\nu}$.

**In section §7** "The Geodesic Line and Equations of Motion of Points", Einstein came back to what he had said in section §2. The worldline of a free material point is a geodesic line. A material point that is free of a gravitational field moves in a straight line and uniformly according to equation (1), its integral curves are geodesics. A material point in gravitational fields moves on a geodesic line in a four-dimensional manifold.[118]

The variational calculation leads to the equation of a geodesic line:[119]

$$(23a) \sum_{\mu} g_{\sigma\mu} \frac{d^2 x_\mu}{ds^2} + \sum_{\mu\nu} \begin{bmatrix} \mu\nu \\ \sigma \end{bmatrix} \frac{dx_\mu}{ds} \frac{dx_\nu}{ds} = 0,$$

where, [120]

$$(\mathbf{24}) \begin{bmatrix} \mu\nu \\ \sigma \end{bmatrix} = \frac{1}{2} \left( \frac{\partial g_{\mu\sigma}}{\partial x_\nu} + \frac{\partial g_{\nu\sigma}}{\partial x_\mu} - \frac{\partial g_{\mu\nu}}{\partial x_\sigma} \right),$$

(24) are the Christoffel symbols of the first kind, derivatives of the metric tensor with respect to the coordinates. Using the Christoffel symbols of the second kind, (24) is, [121]

$$(\mathbf{24a}) \begin{Bmatrix} \mu\nu \\ \tau \end{Bmatrix} = \Sigma_\sigma g^{\sigma\tau} \begin{bmatrix} \mu\nu \\ \sigma \end{bmatrix}$$

Using (24a), equation (23a) can be written in the following way:

$$(\mathbf{23b}) \frac{d^2 x_\tau}{ds^2} + \sum_{\mu\nu} \begin{Bmatrix} \mu\nu \\ \tau \end{Bmatrix} \frac{dx_\mu}{ds} \frac{dx_\nu}{ds} = 0$$

This is the equation of the geodesic line in its most comprehensive form. This equation is equivalent to the equation of motion of a material point in its Minkowski form, where s is the local time. [122]

According to equation (23b), when we are in a flat space-time, the components of the metric tensor are constant, and their derivatives are equal to zero. In this case equation (23b) is reduced to the first term on the left-hand side, which is very similar to the Newtonian equation of motion for a free particle. In fact this is the law of inertia. The second derivative of the coordinate with respect to time is the acceleration component in the direction of this coordinate. When the acceleration is constant the body moves in constant velocity and in a straight line.

Once we move to four-dimensional space, the derivatives of the metric tensor components are no more equal to zero, and thus the Christoffel symbols are not equal to zero. The geodesic line is not anymore straight.

**In section §8** Einstein supplied some more mathematical tools, and in the next section he established the generally covariant equations of physical processes in a gravitational field, equations that correspond to the energy-momentum theorem. He showed that these equations are formally characterized by the divergence of a tensor of rank two $A^{\mu\nu}$ being equal to a covariant four vector $\sum_\mu A^\mu g_{\mu\sigma} = A_\sigma$.

Einstein first obtained the equation,

$$(41a)\ A_\sigma = \frac{1}{\sqrt{g}}\left(\sum_\nu \frac{\partial\left(A_\sigma^\nu \sqrt{g}\right)}{\partial x_\nu} - \frac{1}{2}\sum_{\mu\nu\tau} g^{\tau\mu}\frac{\partial g_{\mu\nu}}{\partial x_\sigma} A_\tau^\nu \sqrt{g}\right)$$

where, $\sum_\mu g_{\sigma\mu} A^{\mu\nu} = A_\sigma^\nu$ is a mixed tensor.[123]

Rewritting equation (41a) using tensor densities (what Einstein called "V-tensors") he obtained:[124]

$$(41b)\ \mathfrak{A}_\sigma = \sum_\nu \frac{\partial \mathfrak{A}_\sigma^\nu}{\partial x_\nu} - \frac{1}{2}\sum_{\mu\tau\nu} g^{\tau\mu}\frac{\partial g_{\mu\nu}}{\partial x_\sigma} \mathfrak{A}_\sigma^\nu,$$

where,

$$A_\sigma \sqrt{g} = \mathfrak{A}_\sigma,\ A_\sigma^\nu \sqrt{g} = \mathfrak{A}_\mu^\nu.$$

Einstein was confused with different kind of tensors: covariant, contravariant, tensor densities – and he said that a choice should be made.[125] He wrote in a footnote that any tensor can be obtained from another tensor of another character by multiplying it with the fundamental tensor, $g_{\mu\nu}$, or with $\sqrt{-g}$.[126]

### 1.10.3 C. "The Equations of Physical Processes in a Given Gravitational Field"

**In section §9**, Einstein was ready to derive the equations of the physical processes in a given gravitational field. Einstein started with "Energy-Momentum Theorem for 'Material Processes'". The generalized equations are formally characterized by the divergence of a tensor of rank two being equal to a four vector. The symmetric energy tensor $T_{\sigma\nu}$ is represented by the mixed tensor density $\mathfrak{T}_\sigma^\nu$, which is the energy tensor of flowing matter [Energietensor der ponderabeln Massenströmung], and the four vector $K_\sigma$ is represented by the covariant four-vector $\overset{\nu}{\underset{\sigma}{\mathfrak{K}}}$ ($K_\sigma = f_x, f_y, f_z, iw$, where, $f$ is the force vector per unit volume externally acting upon the system, and $w$ is the energy per unit volume and unit time supplied to the system).[127]

The divergence is then formed according to equation (41b), and the generally covariant equations that represent energy-momentum conservation are:[128]

$$(42a) \sum_{\nu} \frac{\partial \mathfrak{T}_{\tau}^{\nu}}{\partial x_{\nu}} = \frac{1}{2} \sum_{\mu\tau\nu} g^{\tau\mu} \frac{\partial g_{\mu\nu}}{\partial x_{\sigma}} \mathfrak{T}_{\tau}^{\nu} + \mathfrak{K}_{\sigma}.$$

These equations transform into the original equations of energy-momentum conservation law for the special theory of relativity when:

$g_{\mu\nu}$ = diag (– 1, – 1, – 1, 1).

If there is no gravitational field then $g_{\mu\nu}$ = const., no external forces acting and $\mathfrak{K}_{\sigma} = 0$. Thus,

$$(42b) \sum_{\nu} \frac{\partial \mathfrak{T}_{\sigma}^{\nu}}{\partial x_{\nu}} = 0.$$ [129]

When the gravitational field exists, i.e., when the $g_{\mu\nu}$ are not constant and for vanishing $\mathfrak{K}_{\sigma}$, things are getting much more complicated; the first term in equation (42a) does not vanish. Therefore, one has to demand that for the material system and associated gravitational field combined, the constancy of the total momentum and total energy of matter plus gravitational field should be expressed. Thus a complex quantity designated by $t_{\sigma}^{\nu}$ should exist for the gravitational field such that the following equations apply: [130]

$$(42c) \sum_{\nu} \frac{\partial (\mathfrak{T}_{\sigma}^{\nu} + t_{\sigma}^{\nu})}{\partial x_{\nu}} = 0.$$

In this case of equation (42a) Einstein realizes the importance of the action of the gravitational field on material processes,[131]

$$(46)\ \Gamma_{\nu\sigma}^{\tau} = \frac{1}{2} \sum_{\mu} g^{\tau\mu} \frac{\partial g_{\mu\nu}}{\partial x_{\sigma}},$$

and chooses this equation to be the "components of the gravitational field".

**In section §10** Einstein dealt with "Equations of Motion of Continuously Distributed Masses". Einstein wrote that one can form a mixed tensor density of the form: [132]

$$(48)\ \mathfrak{T}_\sigma^\nu = \rho_0\sqrt{-g}\frac{dx_\nu}{ds}\sum_\mu g_{\sigma\mu}\frac{dx_\mu}{ds},$$

where, the velocity = $dx_\mu/ds$ is a contravariant four-vector and $\rho_0$ is the scalar density of continuously distributed matter.

### 1.10.4 D. "The Differential Laws of the Gravitational Field"

Einstein finally arrived at the then problematic part of his theory, the derivation of the field equations. He started from the coefficients $g_{\mu\nu}$ as functions of the $x_\nu$, the components of the gravitational potential. He requested that the differential equations are satisfied by these quantities. The theory that Einstein has developed in the preceding sections complies with the principle of relativity. From the formal aspects its equations are general, that is, covariant under arbitrary substitutions of the $x_\nu$. Einstein then wrote, "It appears that the necessary demand thereafter is that the differential laws of the $g_{\mu\nu}$ must be generally-covariant. However, we want to show that we have to restrict this demand if we want to fully satisfy the theorem of causality. We shall prove, namely, that it can be impossible that the laws that determine the course of events in a gravitational field are generally covariant".[133] Subsequently, Einstein presented in section §12 the "Proof of Necessary of a Restriction of the Choice of Coordinates", an improved and elaborated version of a Hole Argument:[134]

"We consider a finite portion Σ of the continuum, in which a material process does not occur. The physical events in Σ are completely determined if the quantities $g_{\mu\nu}$ are given as functions of the $x_\nu$ with respect to the coordinate system K used for the description. The totality of these functions will be symbolically denoted by G(x). Let there be a new coordinate system K', outside of Σ, which coincides with K, but within Σ, it diverges from K such that the $g'_{\mu\nu}$ referred to K' as well as the $g_{\mu\nu}$ (including their derivations) are everywhere continuous. The totality of the $g'_{\mu\nu}$ is symbolically denoted by G'(x'). G'(x') and G(x) describe the same gravitational field. If we replace the coordinates $x'_\nu$ by the coordinates $x_\nu$ in the functions $g'_{\mu\nu}$, i.e., if we form G'(x), then G'(x) also describes a gravitational field with respect to K, which however, does not correspond to the actual (i.e., the originally given) gravitational field.

If we assume now the differential equations of the gravitational field are generally covariant, then they are satisfied for G'(x') (with respect to K') if they are satisfied by G(x) with respect to K. They are then also satisfied with respect to K by G'(x). With respect to K there then existed different solutions G(x) and G'(x), which are different from one another, nevertheless at the boundary of the region both solutions coincide, *i.e., what is happening cannot be determined uniquely by generally-covariant differential equations for the gravitational field.*

When we therefore request that the course of events in the gravitational field be completely determined by the laws to be set up, then we are forced to restrict the choice of the coordinate system such as, without a violation of the restrictive conditions, it is impossible to introduce a new coordinate system K' of the type previously characterized . The continuation of the coordinate system into the interior of the region Σ must not be arbitrary".

After presenting the Hole Argument, Einstein presumed he persuaded the reader, and he moved on to **section §13**. Einstein first started the section by saying, "Once we have seen that the coordinate system has to be subject to conditions, we must focus upon several kinds of specializations in the choice of coordinates. A very high degree of specialization is obtained by admitting only linear transformations".[135]

Einstein explained that if one demanded that the equations of physics be covariant merely with respect to *linear* transformations then the theory would lose its main support. The problem is transformation to an accelerated or rotating system. It would no longer be admissible in a theory which is covariant with respect to *linear* transformations only. And thus Einstein's theory would not be able to interpret the physical equivalence of a "centrifugal field" and gravitational field – which Einstein emphasized it in section §1.[136]

Einstein was under the impression that the metric field describing space and time for a rotating system was a solution of the field equations of the "Entwurf" theory. He therefore thought that he managed to fulfill the principle of equivalence he had presented in §1: the centrifugal force, which acts under given conditions upon a body, is determined by precisely the same natural constant as the effect of the gravitational field, such that we have no means to distinguish a centrifugal field from a gravitational field.

Einstein next started with the "Hamiltonian" function H of the contravariant fundamental tensor $g^{\mu\nu}$ and its first derivatives $\partial g^{\mu\nu}/\partial x_\sigma$, where the latter are called $g_\sigma^{\mu\nu}$ for short. Einstein wrote the following equation for the integral J; the integral is extended over a finite part Σ of the continuum:

$$(61)\ J = \int H\sqrt{-g}\, d\tau.$$

The coordinate system Einstein first used was $K_1$. He then asked for the change ΔJ of J when we go from the system $K_1$ to another system $K_2$, which is infinitesimally different from $K_1$.[137]

In section §6 Einstein wrote the following relation for the fundamental tensor $g^{\mu\nu}$:

$$(17a)\ \sqrt{-g}d\tau = d\tau_0.$$ [138]

Einstein considered the change Δ due to infinitesimal transformation of a certain quantity at some point of the continuum, and found according to this relation:

$$(62)\ \Delta\sqrt{-g}d\tau = 0$$

Einstein obtained an expression for ΔH in terms of $\Delta g^{\mu\nu}$ and $\Delta g_\sigma^{\mu\nu}$.
By taking the following relations into account:

$\Delta g^{\mu\nu} = g^{\mu\nu'} - g^{\mu\nu}$,

$\Delta x_\mu = x'_\mu - x_\mu$,.

and using (8), (from section §4), Einstein obtained an expression for $\Delta g^{\mu\nu}$ in terms of $\Delta x_\mu$,

$$(63)\ \Delta g^{\mu\nu} = \sum_\alpha \left( g^{\mu\alpha} \frac{\partial \Delta x_\nu}{\partial x_\alpha} + g^{\nu\alpha} \frac{\partial \Delta x_\mu}{\partial x_\alpha} \right),$$

and an expression for $\Delta g_\sigma^{\mu\nu}$.

Einstein assumed that H depends on $g^{\mu\nu}$ and $g_\sigma^{\mu\nu}$, and that it is invariant under linear transformations. Therefore, ΔH vanished when,

$$\frac{\partial^2 \Delta x_\mu}{\partial x_\alpha \partial x_\sigma} = 0.$$ [139]

Under this assumption he obtained the following expression:

$$(64)\ \frac{1}{2}\Delta H = \sum_{\mu\nu\tau\alpha} g^{\nu\alpha} \frac{\partial H}{\partial g_\sigma^{\mu\nu}} \frac{\partial^2 \Delta x_\mu}{\partial x_\sigma\, \partial x_\alpha}.$$

From this by means of equation (62) Einstein obtained an expression for ΔJ:

$$\frac{1}{2}\Delta J = \int d\tau \sum_{\mu\nu\sigma\alpha} \frac{\partial H\sqrt{-g}}{\partial g_{\sigma}^{\mu\nu}} \frac{\partial^2 \Delta x_\mu}{\partial x_\sigma\, \partial x_\alpha}.$$

And from this by partial integration:

$$(65)\frac{1}{2}\Delta J = \int d\tau \sum_{\mu} (\Delta x_\mu B_\mu) + F.$$

Before Einstein left Zurich he wrote a second paper with Grossmann in 1914, and there he and Grossmann wrote, that the gravitational equations, "*Are covariant with respect to all admissible transformations of the coordinate systems, i.e., with respect to all transformations between coordinate systems which satisfy the [coordinate]* conditions Bσ [...] = 0".[140]

These four conditions were to hold in all those systems, which were adapted coordinate systems, and in which the Einstein-Grossman 1914 field equations were valid. Einstein thus posed a coordinate condition on the field equations and the coordinate systems satisfying this condition were the adapted coordinate systems for the gravitational field. Coordinate transformations between two such adapted coordinate systems were arbitrary non-linear transformations.

Einstein and Grossman restricted the coordinate systems by the conditions $B_\sigma = 0$. And they showed that these conditions were direct consequence of the gravitational equations, but the covariance of the equations was nevertheless far-reaching in these coordinate systems.[141]

In the 1914 review article Einstein restricted the coordinate systems to the adapted coordinate sustem (Angepaßte Koordinatensysteme), using the same conditions: [142]

$$(65a)\ B_\mu = \sum\nolimits_{\alpha\sigma\nu} \frac{\partial^2}{\partial x_\sigma\, \partial x_\alpha}\left(g^{\nu\alpha}\frac{\partial H\sqrt{-g}}{\partial g_{\sigma}^{\mu\nu}}\right).$$

Einstein also wrote an expression for F. F is a surface integral and vanishes if $\Delta x_\mu$ and $\frac{\partial \Delta x_\mu}{\partial \Delta x_\alpha}$ vanish at the boundary.

He considered the finite portion Σ and the coordinate system K. He imagined a series of infinitely related coordinate systems K', K" and so on, so that $\Delta x_\mu$ and $\frac{\partial \Delta x_\mu}{\partial \Delta x_\alpha}$ vanish at the boundaries. For every infinitesimal coordinate transformation between neighboring coordinate systems of the total coordinate systems K, K', K",… we thus obtain: F = 0, and the equation for ΔJ becomes :

$$(66)\ \frac{1}{2}\Delta J = \sum_{\mu} \int d\tau\, \Delta x_{\mu} B_{\mu}.$$

Einstein now chose coordinate systems which are "adapted to the gravitational field".[143] The equations (65a) become:

$$(67)\ B_{\mu} = 0$$

and hold for these adapted systems.

This is a sufficient condition that the coordinate system is adapted to the gravitational field.[144]

Einstein explained that he managed to avoid the difficulty mentioned at the beginning of section §13: he killed two birds with one stone; the physical equivalence between the inertial field and the gravitational field and the rotation metric (so he thought) was a solution of his "Entwurf" field equations. Einstein concluded, "it is indeed the restriction to adapted coordinate systems that did not allow a coordinate system that is given for the exterior of Σ to continue into the interior of Σ in an arbitrary manner".[145]

**In section §14** Einstein proved that,

$$\frac{\mathfrak{G}_{\mu\nu}}{\sqrt{-g}},$$

Einstein's 1914 gravitation tensor, "under limitation to adapted coordinate systems, and substitutions between them, is a covariant tensor.

**Proof that the gravitational tensor $\mathfrak{G}_{\mu\nu}$ is a covariant tensor**

In section §14 of his 1914 review paper, Einstein supplied a proof of a theorem that formed the formal basis for the claim that, if the coordinate system is an adapted coordinate system, then the gravitational tensor $\mathfrak{G}_{\mu\nu}$ is a covariant tensor.

Einstein wrote that equation (65):

$$(65)\frac{1}{2}\Delta J = \int d\tau \sum_{\mu} (\Delta x_{\mu} B_{\mu}) + F$$

leads to a theorem of fundamental importance to the entire theory. If the fundamental tensor $g_{\mu\nu}$ is varied by an infinitely small amount, so that $g_{\mu\nu}$ are replaced by $g^{\mu\nu} + \delta g^{\mu\nu}$, where the $\delta g^{\mu\nu}$ shall vanish at the boundaries of Σ, then H becomes H+δH and the action J becomes J + δJ.

Einstein then claimed that the equation:

(68) $\Delta\{\delta J\} = 0$

always holds whichever way the $\delta g_{\mu\nu}$ might be chosen, provided the coordinate systems ($K_1$ and $K_2$) are *adapted* coordinate systems with respect to the unvaried gravitational field. This means that under the restriction to adapted coordinate systems δJ is an invariant.[146]

In order to prove this Einstein imagined the variations $\delta g^{\mu\nu}$ to be composed of two parts. He thus wrote:

(69) $\delta g^{\mu\nu} = \delta_1 g^{\mu\nu} + \delta_2 g^{\mu\nu}$

$\delta_1 g^{\mu\nu}$ – these are taken in a manner so that the coordinate system $K_1$ is not only adapted to the gravitational field of the $g_{\mu\nu}$, but also to the *varied* gravitational field:

$g^{\mu\nu} + \delta g^{\mu\nu}$.

Therefore,

(67) $B_\mu = 0$

and

(70) $\delta_1 B_\mu = 0$.

$\delta_2 g^{\mu\nu}$ – these are taken without changing the gravitational field, by variation of the coordinate system, variation in the sub-domain of Σ in which the $\delta g^{\mu\nu}$ are not 0, and thus $\delta_2 B_\mu \neq 0$.

The superposition of the two variations is determined by 10 mutually independent functions, and thus – so believed Einstein – will be equivalent to an *arbitrary* variation of the $\delta g^{\mu\nu}$. Hence Einstein reasoned that the proof of his theorem would be completed once equation (68) was proven for the two partial variations.[147]

Einstein proved his theorem and deduced the existence of 10 components, which has tensorial character if we limit ourselves to adapted coordinate systems.[148]

Einstein next came back to equation (61) $J = \int H\sqrt{-g}\, d\tau$. After varying infinitesimally the $g^{\mu\nu}$ he rewrote this equation in the following form:

$$[(61a)]\ dJ = \delta\left\{\int H\sqrt{-g}\, d\tau\right\}$$
$$= \int d\tau \sum_{\mu\nu\sigma} \left\{ \frac{\partial(H\sqrt{-g})}{\partial g^{\mu\nu}} \delta g^{\mu\nu} + \frac{\partial(H\sqrt{-g})}{\partial g_\sigma^{\mu\nu}} \delta g_\sigma^{\mu\nu} \right\},$$

and since: $\delta g_\sigma^{\mu\nu} = \frac{\partial}{\partial x_\sigma}(\delta g^{\mu\nu})$

after partial integration and considering the vanishing of $\delta g^{\mu\nu}$ at the boundary of $\Sigma$,

$$(71)\ \delta J = \int d\tau \sum_{\mu\nu} \delta g^{\mu\nu} \left\{ \frac{\partial H\sqrt{-g}}{\partial g^{\mu\nu}} - \sum_\sigma \frac{\partial}{\partial x_\sigma}\left(\frac{\partial H\sqrt{-g}}{\partial g_\sigma^{\mu\nu}}\right) \right\}.$$

Via his above theorem Einstein proved that under limitation to adapted coordinate systems $\delta J$ was invariant.[149] Using his theorem thus proved, Einstein concluded that since $\delta g^{\mu\nu}$ differs from zero only in infinitely small areas, and since $\sqrt{-g}d\tau$ is a scalar, the integral (the quantity) divided by $\sqrt{-g}$ is also an invariant: [150]

$$(72)\ \frac{1}{\sqrt{-g}} \sum \delta g^{\mu\nu} \mathfrak{G}_{\mu\nu}$$

The quantity in the brackets of (71) and in (72) is:

$$(73)\ \mathfrak{G}_{\mu\nu} = \frac{\partial H\sqrt{-g}}{\partial g^{\mu\nu}} - \sum_\sigma \frac{\partial}{\partial x_\sigma}\left(\frac{\partial H\sqrt{-g}}{\partial g_\sigma^{\mu\nu}}\right),$$

This tensor $\mathfrak{G}_{\mu\nu}$ was going to be Einstein's 1914 gravitation tensor. And thus,

$$(71)\ \delta J = \int d\tau \sum_{\mu\nu} \delta g^{\mu\nu} \mathfrak{G}_{\mu\nu}.$$

The $\mathfrak{G}_{\mu\nu}$ is a covariant tensor density.[151]

Einstein wrote, "It follows that

$$\frac{\mathfrak{G}_{\mu\nu}}{\sqrt{-g}}$$

under limitation to adapted coordinate systems, and substitutions between them, is a covariant tensor, and $\mathfrak{G}_{\mu\nu}$ *itself is the corresponding covariant V-tensor[tensor density] and according to (73) a symmetric tensor*".[152]

**In section §15** Einstein wrote that the tensor $\mathfrak{G}_{\mu\nu}$ has a fundamental role in the field equations of gravitation, and those equations take the place of Poisson's equations. Einstein demanded that the desired field equations would be covariant only with respect to adapted coordinate systems. The proof from the previous section was of crucial importance to the field equations: Einstein could now look for equations that would represent a correlation between the tensors $\mathfrak{G}_{\mu\nu}$ and $\mathfrak{T}_{\tau}^{\nu}$.

Einstein was searching for equations of the form:

$$(74)\ \mathfrak{G}_{\sigma\tau} = \chi \mathfrak{T}_{\sigma\tau}$$

Since the divergence of the energy tensor $\mathfrak{T}_{\tau}^{\nu}$ of the material processes vanishes, according to the above equation, the tensor $\mathfrak{G}_{\sigma\tau}$ or the mixed tensor density $\mathfrak{G}_{\sigma}^{\nu}$ also vanishes. The ten above equations are used to determine the ten functions $g^{\mu\nu}$ if the $\mathfrak{T}_{\sigma\tau}$ are given. And the $g^{\mu\nu}$ must also satisfy the four equations $B_{\mu} = 0$, because the coordinate system is an adapted one. And the equations for the energy tensor:[153]

$$(75)\ \mathfrak{T}_{\sigma\tau} = \textstyle\sum_{\nu} g_{\nu\tau} \mathfrak{T}_{\sigma}^{\nu}$$

$$\mathfrak{T}_{\sigma}^{\nu} = \sum_{\tau} \mathfrak{T}_{\sigma\tau} g^{\nu\tau}.$$

Einstein now went on to determine the function H:

**First,** H depends only on the $g^{\mu\nu}$ and $g_{\sigma}^{\mu\nu}$.

**Second,** it is a scalar with respect to linear transformations.

**Third,** since $\mathfrak{T}_\tau^\nu$ and $\mathfrak{G}_\sigma^\nu$ vanish, by (41b) every gravitational field must satisfy the conservation law:

$$\sum_{\nu\tau} \frac{\partial}{\partial x_\nu}(g^{\tau\nu}\mathfrak{G}_{\sigma\tau}) + \frac{1}{2}\sum_{\mu\nu} \frac{\partial g^{\mu\nu}}{\partial x_\sigma}\mathfrak{G}_{\mu\nu} = 0.$$

Einstein took the coordinate conditions (65a), and equation (73) and wrote the above relation in a succinct form:

$$(76)\ \sum_\nu \frac{\partial S_\sigma^\nu}{\partial x_\nu} - B_\sigma = 0,$$

where,

$$(76a)\ S_\sigma^\nu = \sum_{\mu\tau}\left(g^{\nu\tau}\frac{\partial H\sqrt{-g}}{\partial g^{\sigma\tau}} + g_\mu^{\nu\tau}\frac{\partial H\sqrt{-g}}{\partial g_\mu^{\sigma\tau}} + \frac{1}{2}\delta_\sigma^\nu H\sqrt{-g} - \frac{1}{2}g_\sigma^{\mu\tau}\frac{\partial H\sqrt{-g}}{\partial g_\nu^{\mu\tau}}\right).$$

The ten field equations determine the ten functions $g^{\mu\nu}$, but the $g^{\mu\nu}$ must also satisfy the four equations $B_\mu = 0$; "so we have more equations than we have functions to be found". This is only possible if:

$$(77)\ S_\sigma^\nu \equiv 0.$$

**Fourth,** this condition was not enough to fully determine H. Einstein thus "without specifying a formal reason for the demand", requested furthermore that H would be a homogeneous function of the second degree in the $g_\sigma^{\mu\nu}$.

In this case H is completely determined up to a constant factor. Einstein found H to be a linear combination of five quantities. Einstein finally managed to equate H – aside from a constant factor – to the fourth one of these quantities:

$$(78)\ H = \frac{1}{4}\sum_{\alpha\beta\tau\rho} g^{\alpha\beta}\frac{\partial g_{\tau\rho}}{\partial x_\alpha}\frac{\partial g^{\tau\rho}}{\partial x_\beta}.$$

This choice of H satisfies condition (77) for $S_\sigma^\nu$.[154]

Einstein concluded, "We have now arrived in a purely formal way, i.e., without direct recourse to our physical knowledge of gravitation, at quite specific field equations".[155]

In fact Einstein's derivation was not completely formal. Einstein would later discover that his choice of H was also driven by physical premises.

Einstein next wrote the field equations themselves in explicit notation. Multiplying (74) by $g^{\nu\tau}$ and summing over the index τ, and inserting the value for $\mathfrak{G}_{\mu\nu}$ from equation (73) Einstein obtained:

$$(80)\ \chi\mathfrak{T}_\sigma^\nu = \sum_{\tau\alpha} g^{\nu\tau}\left(\frac{\partial H\sqrt{-g}}{\partial g^{\sigma\tau}} - \frac{\partial}{\partial x_\alpha}\left[\frac{\partial H\sqrt{-g}}{\partial g_\alpha^{\sigma\tau}}\right]\right),$$

or,

$$(80a) - \sum_{\alpha\tau}\frac{\partial}{\partial x_\alpha}\left(g^{\nu\tau}\frac{\partial H\sqrt{-g}}{\partial g_\alpha^{\sigma\tau}}\right) = \chi\mathfrak{T}_\sigma^\nu + \sum_{\alpha\tau}\left(-g^{\nu\tau}\frac{\partial H\sqrt{-g}}{\partial g^{\sigma\tau}} - g_\alpha^{\nu\tau}\frac{\partial H\sqrt{-g}}{\partial g_\alpha^{\sigma\tau}}\right).$$

Since the coordinate system is an adapted one ($B_\mu = 0$), the conservation law holds:

$$(80b)\ \sum_\nu \frac{\partial}{\partial x_\nu}\left\{\mathfrak{T}_\sigma^\nu + \frac{1}{\chi}\sum_{\alpha\tau}\left(-g^{\nu\tau}\frac{\partial H\sqrt{-g}}{\partial g^{\sigma\tau}} - g_\alpha^{\nu\tau}\frac{\partial H\sqrt{-g}}{\partial g_\alpha^{\sigma\tau}}\right)\right\} = 0.$$

Einstein then returned to the value (78) for H that he has just obtained above. He also defined in section §10 the naturally measured quantities dξ (and the coordinate system in which the dξ are referred to in the infinitesimally small, the "normal system"),

$$ds^2 = \sum_{\mu\nu} g_{\mu\nu}dx_\mu dx_\nu = -d\xi_1^2 - d\xi_2^2 - d\xi_3^2 - d\xi_4^2.$$

With the equation for H (and an additional equation), and with the above equation he rewrote the field equations and the conservation law in the following form:

$$(81) \sum_{\alpha\beta}\frac{\partial}{\partial x^\alpha}\left(\sqrt{-g}\, g^{\alpha\beta}\Gamma_{\sigma\beta}^\nu\right) = -\chi(\mathfrak{T}_\sigma^\nu + t_\sigma^\nu),$$

where,

$$(81a)\ \Gamma_{\sigma\beta}^\nu = \frac{1}{2}\sum_\tau g^{\nu\tau}\frac{\partial g_{\sigma\tau}}{\partial x_\beta},$$

are equations (46): what Einstein called the components of the field strength of the gravitational field, and:

$$(81b)\ t_\sigma^\nu = \frac{\sqrt{-g}}{\chi} \sum_{\mu\rho\tau\tau'} \left( g^{\nu\tau}\Gamma^\rho_{\mu\sigma}\Gamma^\mu_{\rho\tau} - \frac{1}{2}\delta^\nu_\sigma g^{\tau\tau'}\Gamma^\rho_{\mu\tau}\Gamma^\mu_{\rho\tau'} \right),$$

And,

$$(42c) \sum_\nu \frac{\partial}{\partial x_\nu}(\mathfrak{T}_\sigma^\nu + t_\sigma^\nu) = 0.$$

These were the differential equations of the gravitational field.[156]

Einstein emphasized that the $t_\sigma^\nu$ do not have tensorial covariance with respect to arbitrary admissible transformations but only with respect to *linear* transformations. Nevertheless, Einstein called $t_\sigma^\nu$ the energy tensor of the gravitational field. Einstein emphasized an important aspect of the above field equations: "the energy tensor of the gravitational field itself acts in the same way as field-excitation, like the energy tensor of matter.[157]

### 1.11 The 1914 Equations that were Later Replaced in the November 4, 1915 Paper

Einstein obtained the following main schemes and equations in his 1914 review paper that assisted him in developing the equations of the November 4, 1915 paper and in addition were replaced by new equations in this latter paper:

#### 1.11.1. The $\sqrt{g}$ or $\sqrt{-g}$ factor

In section §6 of the 1914 review paper Einstein wrote the following for the fundamental tensor: [158]

$$(10) \sum_\sigma g_{\mu\sigma} g^{\nu\sigma} = \delta_\mu^\nu.$$

where $\delta_\mu^\nu$ is the Kronecker delta defined by:

$$\delta_\mu^\nu = 1 \text{ if } \mu = \nu$$

$$\delta_\mu^\nu = 0 \text{ if } \mu \neq \nu$$

And according to the multiplication theorem for determinants,

$$\left|\sum_\alpha (g_{\mu\alpha} g^{\alpha\nu})\right| = |\delta_\mu^\nu| = 1.$$

From which it follows,

$$(11)\ \ |g_{\mu\nu}| \cdot |g^{\mu\nu}| = 1.$$

Einstein also obtained using (2b) the integral:

$$d\tau_0^* = \int dX_1 dX_2 dX_3 dX_4,$$

which is an invariant – completely independent of the choice of coordinates. Einstein wrote,

$$(13)\ dX_\sigma = \sum_\mu \alpha_{\sigma\mu} dx_{\mu,}$$

by which he gave another expression for the invariant,

$$(14)\ \ d\tau_0^* = |\alpha_{\sigma\mu}| d\tau.$$

If,

$$d\tau = \int dx_1 \ldots dx_4$$

and

$$d\tau_0^* = \int dX_1 \ldots dX_4$$

extended over some elementary area, then according to (2b) and (13)

$$(15)\quad g_{\mu\nu} = \sum_{\sigma} \alpha_{\sigma\mu}\alpha_{\sigma\nu},$$

and according to the multiplication theorem for determinants,

$$(16)\quad |g_{\mu\nu}| = \left|\sum_{\sigma} \alpha_{\sigma\mu}\alpha_{\sigma\nu}\right| = |\alpha_{\mu\nu}|^2$$

Einstein rewrote the equations (14) for $d\tau_0^*$ in the following form:

**(17)** $\sqrt{g}d\tau \ = d\tau_0^*,$

where, $|g_{\mu\nu}| = g$ for short.[159]

According to (2b) the dXσ correspond to the coordinates in special relativity: three of these are real valued and one is imaginary (i.e., the $dX_4$). Consequently dτ is imaginary. However, in special relativity the determinant g with real value time coordinates is negative, because $g_{\mu\nu}$ is diagonal (– 1, – 1, – 1, 1). $\sqrt{g}$ is therefore also imaginary. In order to avoid imaginary coordinates, Einstein put,

$$d\tau_0 = \frac{1}{i}\int dX_1 dX_2 dX_3 dX_4,$$

and instead of equation (17) wrote for dτ the following equation:

**(17a)** $\boldsymbol{\sqrt{-g}d\tau \ = d\tau_0}$.[160]

### 1.11.2. The fundamental tensor of Ricci and Levi-Civita

After equation (17a), under the heading "The anti-symmetric fundamental tensor of Ricci and Levi-Civita" Einstein developed some relations he would not use in the 1914 paper, but would come back to in his papers of November 1915: [161]

$$(19)\ G_{iklm} = \sqrt{g}\delta_{iklm}$$

Einstein wrote that this is a covariant tensor. $\delta_{iklm}$ is the Kronecker delta and it is +1 or –1 depending on the iklmn and their equality to 1234 by an even or odd permutation of the indices.

From the tensor $G_{iklm}$ one can form a contravariant tensor by mixed multiplication according to:

$$(21)\ \sum_{\alpha\beta\lambda\mu} G_{\alpha\beta\lambda\mu} g^{\alpha i} g^{\beta k} g^{\lambda l} g^{\mu m} = G_{iklm.}$$

Using equation (19), the multiplication theorem of determinants, and equation (11) Einstein was led to the result, $\frac{1}{\sqrt{g}}\delta_{iklm}$. With this he proved that,

$$(21a)\ G^{iklm} = \frac{1}{\sqrt{g}}\delta_{iklm}$$

is a contravariant anti-symmetric tensor. [162]

According to (19) and to (21a) $G_{iklm} \neq G^{iklm}$.

Einstein then mentioned the importance of the mixed tensor $G_{ik}^{lm}$ in the theory of anti-symmetric tensors. It is formed from the fundamental tensor $g_{\mu\nu}$, and its components are:

$$(22)\ G_{ik}^{lm} = \sum_{\alpha\beta} \sqrt{g}\delta_{ik\alpha\beta} g^{\alpha l}\, g^{\beta m} = \sum_{\alpha\beta} \frac{1}{\sqrt{g}}\delta_{lm\alpha\beta}\, g_{\alpha i} g_{\beta k}.$$

Einstein said that the tensorial character of the two expressions was evident. But the problem was to prove that the two expressions – the one on the right and the one on the left were equal.

There was a problem with the factors $\sqrt{g}$ and $\frac{1}{\sqrt{g}}$. Einstein managed to replace the $\frac{1}{\sqrt{g}}$ on the right-hand side with a $\sqrt{g}$ by using (21) and (19). He obtained the following formula:

$$\sum_{\lambda\mu\rho\sigma\alpha\beta} \sqrt{g}\delta_{\lambda\mu\rho\sigma} g^{\lambda l} g^{\mu m} g^{\rho\alpha} g^{\sigma\beta} g_{\alpha i} g_{\beta k}$$

And after summation over α and β, and considering (10) he got:

$$\sum_{\lambda\mu} \sqrt{g}\delta_{\lambda\mu ik} g^{\lambda l} g^{\mu m}.$$

This expression, so said Einstein, differs from the left-hand side expression of (22) only in the notation of the summation indices "and in the (irrelevant) sequence of the index-pairs λμ and ik in $\delta_{\lambda\mu ik}$".[163]

Einstein thus concluded that the mixed tensor $G_{ik}^{lm}$ is anti-symmetric in its indices i, k as well as l, m, as can be seen from (22).[164]

### 1.11.3. The Geodesic line

In section §7 Einstein showed that a material point free of a gravitational field moves in a straight line and uniformly according to (1).[165]

A material point in gravitational fields moves on a geodesic line in a four-dimensional manifold.[166] The equation of a geodesic line is (23a), where, (24) are the Christoffel symbols of the first kind. Using the Christoffel symbols of the second kind **(24)** is **(24a)**. Using (24a), equation (23a) is written as **(23b)** the equation of the geodesic line in its most comprehensive form.

Recall that at the end of section §9 Einstein wrote equation (46), the action of the gravitational field on material processes, the "components of the gravitational field" [compare to equation (24)].[167] $\Gamma_{\nu\sigma}^{\tau}$ entered into the "Entwurf" 1914 field equations (81). However, by November 4 Einstein already understood that (46) is a truncated version of (24a), and that he should have used as components of the gravitational field the Christoffel symbols. Einstein thus dismissed equation (46) and preferred equation (24a) upon it.

### 1.11.4. Formula for extension of tensors

In section §8 Einstein wrote the formula, which Christoffel had already written for producing a covariant tensor from rank *l* another one of rank (*l* +1):

$$(29)\ A_{\alpha_1 \dots \alpha_l s} = \frac{\partial A_{\alpha_\tau \dots \alpha_l}}{\partial x_s} - \sum_\tau \left[ \begin{Bmatrix} \alpha_1 s \\ \tau \end{Bmatrix} A_{\tau\alpha_2 \dots \alpha_l} + \begin{Bmatrix} \alpha_2 s \\ \tau \end{Bmatrix} A_{\alpha_1 \tau \alpha_3 \dots \alpha_l} + \cdots \right]$$

Einstein multiplied this equation by the contravariant fundamental tensor and after manipulation arrived at:

$$(30)\ A_s^{\alpha_1 \dots \alpha_l} = \frac{\partial A^{\alpha_1 \dots \alpha_l}}{\partial x_s} + \sum_\tau \left[ \begin{Bmatrix} s\tau \\ \alpha_1 \end{Bmatrix} A^{\tau\alpha_2 \dots \alpha_l} + \begin{Bmatrix} s\tau \\ \alpha_2 \end{Bmatrix} A^{\alpha_1 \tau \alpha_3 \dots \alpha_l} + \cdots \right]$$

This equation is the extension of the contravariant tensor (of rank *l*). This produces a mixed tensor of rank *l* + 1. It can be transformed into a contravariant tensor of rank *l* – 1 by contraction, inner multiplication with the fundamental tensor (10). It can be done in *l* different ways. *l* different divergences of a contravariant tensor are distinguished and Einstein gives one of them:

$$(31)\ A^{\alpha_1 \dots \alpha_{l-1}} = \sum_{\alpha_l s} A_s^{\alpha_1 \dots \alpha_l} \delta_{\alpha_l}^{s}.$$

With a symmetrical and anti-symmetrical tensors, the end result of the divergence is independent of the privileged $\alpha_\nu$.[168]

Subsequently, Einstein used equations (24) and (24a) (and an additional formula he obtained by differentiating $|g_{\mu\nu}| = g$) and arrived at equation (33):[169]

$$(33) \sum_\tau \begin{Bmatrix} \mu\tau \\ \tau \end{Bmatrix} = \sum_\tau \begin{Bmatrix} \tau\mu \\ \tau \end{Bmatrix} = \frac{1}{2} \sum_{\tau\alpha} g^{\tau\alpha} \frac{\partial g_{\tau\alpha}}{\partial x_\mu} = \frac{1}{\sqrt{g}} \frac{\partial \sqrt{g}}{\partial x_\mu}.$$

And using an equation for the extension of a covariant four vector,[170]

$$(28a)\ A_{\mu\nu} = \frac{\partial A_\mu}{\partial x_\nu} - \sum_\tau \begin{Bmatrix} \tau\mu \\ \tau \end{Bmatrix} A_\tau,$$

and from (30) the extension $A_\nu^\mu$ of the covariant for vector $A^\tau$,

$$A_\nu^\mu = \frac{\partial A^\mu}{\partial x^\nu} + \sum_\tau \begin{Bmatrix} \nu\tau \\ \mu \end{Bmatrix} A^\tau$$

And the divergence of this and (33) lead to,[171]

$$(37)\ \Phi = \frac{1}{\sqrt{g}}\sum_{\mu}\frac{\partial}{\partial x_{\mu}}\left(\sqrt{g}A^{\mu}\right).$$

(29) and (30) applied to covariant and contravariant tensors of rank two, the above equations produce tensors of rank three [Einstein hence rewrote equation (30) as (30a) in his 1914 paper.[172]

Equations (29) and (30) produce tensors of rank three,

$$(29a)\ A_{\mu\nu s} = \frac{\partial A_{\mu\nu}}{\partial x_s} - \sum_{\tau}\left[\begin{Bmatrix}\mu s\\ \tau\end{Bmatrix}A_{\tau\nu} + \begin{Bmatrix}\nu s\\ \tau\end{Bmatrix}A_{\mu\tau}\right],$$

$$(30a)\ A_s^{\mu\nu} = \frac{\partial A^{\mu\nu}}{\partial x_s} + \sum_{\tau}\left[\begin{Bmatrix}s\tau\\ \mu\end{Bmatrix}A^{\tau\nu} + \begin{Bmatrix}s\tau\\ \nu\end{Bmatrix}A^{\mu\tau}\right].$$

Einstein then wrote the divergence of $A^{\mu\nu}$ with respect to the index ν, which follows from equation (31), (30a) and (33):

$$(39)\ A^{\mu} = \sum_{s\nu} A_s^{\mu\nu}\delta_\nu^s = \frac{1}{\sqrt{g}}\left(\sum_{\nu}\frac{\partial\left(A^{\mu\nu}\sqrt{g}\right)}{\partial x_\nu} + \sum_{\tau\nu}\begin{Bmatrix}\tau\nu\\ \mu\end{Bmatrix}A^{\tau\nu}\sqrt{g}\right),$$

Einstein formed a covariant four-vector and set it equal to the divergence of $A^{\mu\nu}$. For anti-symmetrical $A^{\mu\nu}$, (39) produces the divergence of $A^{\mu}$:

$$(40)\ A^{\mu} = \frac{1}{\sqrt{g}}\sum_{\nu}\frac{\partial\left(A^{\mu\nu}\sqrt{g}\right)}{\partial x_\nu}$$

Einstein formed a covariant four-vector $\sum_\mu A^\mu g_{\mu\sigma} = A_\sigma$, which was reciprocal to $A^{\mu}$. And if $A^{\mu\nu}$ is symmetrical then the divergence is:

$$(41a)\ A_\sigma = \frac{1}{\sqrt{g}}\left(\sum_{\nu}\frac{\partial\left(A_\sigma^\nu\sqrt{g}\right)}{\partial x_\nu} - \frac{1}{2}\sum_{\mu\nu\tau} g^{\tau\mu}\frac{\partial g_{\mu\nu}}{\partial x_\sigma}A_\tau^\nu\sqrt{g}\right)$$

where, $\sum_\mu g_{\sigma\mu}A^{\mu\nu} = A_\sigma^\nu$ is a mixed tensor.[173]

In the November 4 paper Einstein was going to use the highlighted equations and especially to simplify some of them with his new postulate of determinant 1.

### 1.11.5. The Riemann-Christoffel tensor

On page 1053 Einstein came back to the the Riemann-Christoffel Tensor he had considered in his Zurich Notebook. However, he did not contract it to the Ricci tensor, and he *did not* use the Riemann tensor in his 1914 field equations.[174]

On page 22R of the Zurich Notebook Einstein wrote the heading "Grossmann". He wrote a form of the Ricci tensor in terms of the Christoffel symbols and their derivatives. This was a fully covariant Ricci tensor in a form resulting from contraction of the Riemann tensor:

$$(13)_{\text{November}} \qquad T_{il} = \sum_{\kappa l} \frac{\partial}{\partial x_l}\begin{Bmatrix} i\kappa \\ \kappa \end{Bmatrix} - \frac{\partial}{\partial x_\kappa}\begin{Bmatrix} il \\ \kappa \end{Bmatrix} + \begin{Bmatrix} i\kappa \\ \lambda \end{Bmatrix}\begin{Bmatrix} \lambda l \\ \kappa \end{Bmatrix} - \begin{Bmatrix} il \\ \lambda \end{Bmatrix}\begin{Bmatrix} \lambda\kappa \\ \kappa \end{Bmatrix},$$

Einstein divided the Ricci tensor into two parts – a tensor of second rank and a presumed gravitational tensor:

$$T_{il} = \left(\frac{\partial T_i}{\partial x_l} - \Sigma\begin{Bmatrix} il \\ \lambda \end{Bmatrix} T_\lambda\right) - \Sigma_{\kappa l}\frac{\partial}{\partial x_\kappa}\begin{Bmatrix} il \\ \kappa \end{Bmatrix} - \begin{Bmatrix} i\kappa \\ \lambda \end{Bmatrix}\begin{Bmatrix} l\lambda \\ \kappa \end{Bmatrix}$$

The second term in the above equation which Einstein called "presumed gravitation tensor" was:

$$(13a)_{\text{November}} \qquad R_{ik} = \sum_{\kappa l} \frac{\partial}{\partial x_\kappa}\begin{Bmatrix} il \\ \kappa \end{Bmatrix} - \begin{Bmatrix} i\kappa \\ \lambda \end{Bmatrix}\begin{Bmatrix} l\lambda \\ \kappa \end{Bmatrix}$$

Setting (13a) equal to the energy-momentum tensor $T_{\mu\nu}$, multiplied by the gravitational constant κ, one arrives at the field equations of Einstein's paper of November 4, 1915 [equations (16)].[175]

In his 1914 paper Einstein formed by equation (29) (twofold extension) from the covariant four-vector $A_\mu$ the tensor of rank three ($A_{\mu\nu\lambda}$). This tensor is equal to five terms:[176]

$$A_{\mu\nu\lambda} = \frac{\partial^2 A_\mu}{\partial x_\nu \partial x_\lambda} - \sum_\tau \left[ \begin{Bmatrix} \mu\lambda \\ \tau \end{Bmatrix} \frac{\partial A_\tau}{\partial x_\nu} + \begin{Bmatrix} \mu\nu \\ \tau \end{Bmatrix} \frac{\partial A_\tau}{\partial x_\lambda} \right]$$
$$- \sum_\tau \begin{Bmatrix} \nu\lambda \\ \tau \end{Bmatrix} \frac{\partial A_\mu}{\partial x_\tau} + \sum_{\sigma\tau} \begin{Bmatrix} \nu\lambda \\ \tau \end{Bmatrix} \begin{Bmatrix} \tau\mu \\ \sigma \end{Bmatrix} A_\sigma - \sum_\sigma \left[ \frac{\partial}{\partial x_\lambda} \begin{Bmatrix} \mu\nu \\ \sigma \end{Bmatrix} - \sum_\tau \begin{Bmatrix} \mu\lambda \\ \tau \end{Bmatrix} \begin{Bmatrix} \nu\tau \\ \sigma \end{Bmatrix} \right] A_\sigma .$$

The fifth term in the square brackets is equal to a tensor of rank two times $A_\sigma$, equation $(13a)_{\text{November}}$ above.

From the equation for $A_{\mu\nu\lambda}$ Einstein concluded that $(A_{\mu\lambda\nu} - A_{\mu\nu\lambda})$ is also a covariant tensor of rank three, and consequently he wrote:

$$\sum_\sigma \left[ \frac{\partial}{\partial x_\lambda} \begin{Bmatrix} \mu\nu \\ \sigma \end{Bmatrix} - \frac{\partial}{\partial x_\nu} \begin{Bmatrix} \mu\lambda \\ \sigma \end{Bmatrix} + \sum_\tau \left( \begin{Bmatrix} \mu\nu \\ \tau \end{Bmatrix} \begin{Bmatrix} \lambda\tau \\ \sigma \end{Bmatrix} - \begin{Bmatrix} \mu\lambda \\ \tau \end{Bmatrix} \begin{Bmatrix} \nu\tau \\ \sigma \end{Bmatrix} \right) \right] A_\sigma .$$

The terms in square brackets were a tensor of rank four, the Riemann-Christoffel Tensor:

$$K^\sigma_{\mu\nu\lambda} = \frac{\partial}{\partial x_\lambda} \begin{Bmatrix} \mu\nu \\ \sigma \end{Bmatrix} - \frac{\partial}{\partial x_\nu} \begin{Bmatrix} \mu\lambda \\ \sigma \end{Bmatrix} + \sum_\tau \left( \begin{Bmatrix} \mu\nu \\ \tau \end{Bmatrix} \begin{Bmatrix} \lambda\tau \\ \sigma \end{Bmatrix} - \begin{Bmatrix} \mu\lambda \\ \tau \end{Bmatrix} \begin{Bmatrix} \nu\tau \\ \sigma \end{Bmatrix} \right),$$

which is covariant in μ, ν, λ and contravariant in σ. All components of this tensor vanish when the $g_{\mu\nu}$ are constants.[177] Immediately after Einstein considered this possibility from the mathematical point of view in his 1914 paper, but did not develop it from the physical point of view, he returned to the scalar factor $\sqrt{-g}$ and to the solution of the physical-mathematical problem *of the occurrence of this factor in the equations of his theory*.

**1.11.6. The $\sqrt{g}$ or $\sqrt{-g}$ and the V-tensors**

Recall that one important type of tensors, according to Einstein, is the V-tensors (Volume-tensors), that is, the tensor density. Tensor components frequently occur when multiplied by the factor $\sqrt{-g}$. Therefore, Einstein said, that he wanted a particular name for this kind of tensor, when the tensor components are multiplied by $\sqrt{g}$ or $\sqrt{-g}$:

$$A_\sigma\sqrt{g} = \mathfrak{A}_\sigma,\ A_\sigma^\nu\sqrt{g} = \mathfrak{A}_\mu^\nu.$$

These are, so explained Einstein, V-tensors.

When multiplied by dτ they represent tensors in the previously defined sense, because $\sqrt{g}d\tau$ is a scalar.[178]

*This last fact could also signify to Einstein that his "Entwurf" theory demanded correction.* However, instead of equating the $\sqrt{g}$ or $\sqrt{-g}$ to 1 Einstein chose the complicated way of the V-tensors.

Einstein then rewrote equation (41a) according to the V-tensors notation as (41b).[179] (41b) differs from (41a) by the factors $\frac{1}{\sqrt{g}}$ and $\sqrt{g}$, which appear in the equation. Thus the 1914 law of divergence (41a) was not the same as that for V-tensors **(41b)**. Einstein represented the tensor $T_{\sigma\nu}$ by the mixed V-tensor [tensor density] $\mathfrak{T}_\sigma^\nu$, and the four vector $K_\sigma$ by the covariant V-four-vector $\mathfrak{K}_\sigma^\nu$. The divergence was then formed according to equation (41b), and the generalization of equation (42) gave the generally covariant equations that represented energy-momentum conservation **(42a)**.[180]

Ecall that Einstein claimed that in the general theory of relativity, unlike the "original" (special) theory of relativity, there are tensors of different character: covariant, contravariant, mixed, V-tensors – and a choice should be made.[181] He wrote that any tensor can be obtained from another tensor of another character by multiplying it with the fundamental tensor, $g_{\mu\nu}$, or with $\sqrt{-g}$.[182]

**1.12 Levi-Civita's objection to Einstein's Review Article**

Tullio Levi-Civita from Padua, one of the founders of tensor calculus, carefully read Einstein's paper and attacked Einstein's "Entwurf" theory. He objected to the major problematic element in Einstein's theory, which reflected its global problem: its field equations were restricted to an adapted coordinate system. The major problematic element was Einstein's theorem and its proof from section §14: these supplied the formal basis for Einstein's belief that if the coordinate system was an adapted coordinate system, then the gravitational tensor $\mathfrak{G}_{\mu\nu}$ was a covariant tensor.

In an exchange of letters and postcards that began on March 1915 and ended in May 1915, Levi-Civita presented his objections to Einstein's proof: In his correspondence with Einstein he demonstrated to the latter that neither was δJ invariant, nor was $\frac{\mathfrak{G}_{\mu\nu}}{\sqrt{-g}}$ a covariant tensor for adapted coordinate systems (within a theory which was limited in its covariance).

Einstein tried to find ways to save his proof by answering Levi-Civita quandaries and demonstrations. Einstein was hard to give up his proof, because the proof enabled him to restrict his theory according to the Hole Argument.

Carlo Cattani and Michelangelo De Maria, who wrote in 1989 about the controversy between Einstein and Levi-Civita, claimed that "Levi-Civita's criticisms contributed to stimulating an early growth of Einstein's 'dissatisfaction' with his *Entwurf* theory. […] Levi-Civita did not question directly the limited covariance properties of Einstein's *Entwurf* equations; instead, he shot his mathematical darts against the proof of a theorem crucial to Einstein's variational derivation of the *Entwurf* equations and, in particular, contested the covariance properties of the so-called gravitation tensor. After many fruitless attempts to rebut Levi-Civita's criticism and to find a more convincing proof of that theorem, Einstein was obliged for the first time to admit that both the proof of that theorem and its consequences were not correct".[183]

Unfortunately, all of Levi-Civita's letters but one single letter were lost, but Einstein's letters to Levi-Civita were saved. And this was typical to Einstein. He did not save letters and used to discard them. Only one letter from March 28, 1915 was saved; because that was the letter which Einstein enclosed back on April 2, 1915 in his letter to Levi-Civita "so that I can refer to it without any inconvenience to you".[184]

Nevertheless, one can easily reconstruct the contents of Levi-Civita's letters from Einstein's replies to his letters, because Einstein recapitulates the contents of Levi-Civita's letters, afterwhich he replies to the latter's objections.[185]

## 2 The re-Rise and Fall of the "Entwurf" Theory

### 2.1 June-August 1915: Einstein presents the "Entwurf" Theory in Göttingen

A short while after the correspondence with Levi-Civita, Einstein was invited by David Hilbert to go to Göttingen to present his New ("Entwurf") General Theory of Relativity. Einstein held the lectures under the auspices of the Wolfskehl Foundation in Göttingen the week beginning Monday, 28 June and he returned to Berlin on July 5, 1915.[186]

All that is left from these lectures is 11 pages of notes by an unknown auditor of part of these lectures[187]. In this *Nachschrift* the unknown auditor reported about Einstein's talk on "The gravitational Field and the Equation of Motion of a material Point in this Field". He wrote equations (1) and (2a), which also appeared in Einstein's 1914 paper. According to the report, in his lectures Einstein explained that instead of $x_\nu$ let us introduce other coordinates $x_\nu'$, then:

$$ds^2 = \sum g'_{\mu\nu} dx'_\mu dx'_\nu$$

And the relation between the fundamental tensor $g_{\mu\nu}$ and $g'_{\mu\nu}$ is:

$$g'_{\mu\nu} = \sum_{\alpha\beta} \frac{\partial x_\alpha}{\partial x'_\mu} \frac{\partial x_\beta}{\partial x'_\nu} g_{\alpha\beta},$$

The fundamental tensor being covariant and transforms according to the above equation, where,

$$\sum g_{\mu\nu} dx_\mu dx_\nu = \sum g'_{\mu\nu} dx'_\mu dx'_\nu$$

results.[188]

Any gravitataional field is characterized by 10 $g_{\mu\nu}$ functions, such that when (2a) is the equation of motion of the material point in this gravitational field, equation (1) is satisfied.[189]

After this demonstration the unknown auditor wrote the title "Methods of Measurement" and he reported about Einstein's measurable coordinates in the infinitely small region, equation (2b) from Einstein's 1914 paper; this equation also appeared in the report of the unknown auditor.

By this ended the report of Einstein's Wolfskehl Göttingen lectures. The report did not include Einstein's "Entwurf" field equations, and neither did it include Einstein's theorem and its proof from section §14 of his 1914 paper; the problematic issue which Levi-Civita could not accept. Thus we do not know whether Einstein had adopted some of Levi-Civita's suggestions from their spring correspondence and included them in his Wolfskehl Göttingen lectures. Perhaps Einstein did not even mention the theorem and proof from section §14 in the Wolfskehl lectures? There is no documentation on this issue.

In Göttingen Einstein met Hilbert. On July 7 Einstein wrote Zangger about his impression of Hilbert; "I got to know and love him. I held six two-hour lectures there on the gravitation theory, which is now clarified very much, and had the experience of convincing the mathematician friends there completely".[190]

Einstein felt he was able to convince Hilbert, this as opposed to the mathematician Levi-Civita who was not convinced by Einstein's proof from section §14 of his 1914 paper. However, Tensor Calculus – having its origin in Gauss, Riemann and Christoffell's differentiall geometry – was developed initially by Ricci and his pupil Levi-Civita. They have published their well-known memoir inaugurating the topic. When Grossmann helped Einstein in his search for a gravitational tensor, he referred him to their works.[191] Thus Levi-Civita finding in spring 1915 a grave mathematical problem in Einstein's gravitational tensor for adapted coordinate systems should have alerted Einstein. But it seems that Einstein still hoped he could save his "Entwurf" field equations.

A week after Einstein wrote Zangger he wrote Sommerfeld, on July 15th. Einstein wrote about the book the latter edited: Lorentz, Hendryk, A., Einstein, Albert, and Minkowski, Herman, *Das Relativitätsprinzip*, published by Druck und Verlag von B. G. Teubner in Leipzig und Berlin.

The book first appeared in 1913 and included the major papers related to the "original" special theory of relativity. Sommerfeld intended to publish a second edition and to include Einstein's latest version of the 1914 general theory of relativity. Einstein wrote to Sommerfeld that he preferred seeing this volume "printed unchanged, without including the general theory of relativity, because none of the current presentations on the latter is complete. Easiest would be to use the first of the Annalen-papers[192] and the Academy-paper [the "Entwurf" 1914 paper].[193] However, I intend to write a special little book as an introduction to relativity theory, with the essential aim of treating from the very beginning the general theory of rel".[194] In July 1915, Einstein could be ambivalent in his attitude toward the 1914 "Entwurf" theory: he was not committed to it as before, but continued to maintain it and intended to write a special little book as an introduction to this theory.

Two or three weeks later, Einstein wrote Zangger again, "I have probably written you that I held 6 lectures at Göttingen, where I was able to convince Hilbert of the general theory of relativity. I am enchanted with the latter, a man of admirable energy and independent personality in all things. Sommerfeld is also beginning to agree; Planck and Laue stay aloof".[195]

Meanwhile, Einstein wrote Wander and Geertruida de Hass[196] and told them: "at the end of June-beginning of July I held six detailed lectures at Göttingen on general relativity theory. To my great satisfaction I succeeded in convincing Hilbert and [Felix] Klein completely".[197]

The "Entwurf" empire was a little re-rising for a couple of summer weeks just to fall afterwards. Einstein went for an excursion to Zurich. The problems that had already been discovered initially in the spring correspondence with Levi-Civita surfaced after Einstein came back from a trip to Switzerland in September 1915.

## 2.2 The Three Problems that Led to the Fall of the "Entwurf" Theory

Einstein told David Hilbert on November 7, 1915, "after I realized about 4 weeks ago that my method of proof was fallacious".[198] He was thus very likely referring to the beginning of October 1915. The final decision to give up the "Entwurf" field equations probably occurred to Einstein at the end of September- the beginning of October 1915. Einstein added in the same letter to Hilbert that his colleague Sommerfeld wrote him that Hilbert had objected to the 1914 "Entwurf" foundations paper. [199] Einstein believed up till then that he managed to convince Hilbert completely, but then sometime in October everything had shattered. Hilbert, exactly like Levi-Civita, objected to the 1914 "Entwurf" foundations paper.

By October 9 of that year Paul Hertz challenged Einstein's Hole Argument from section §12 of his 1914 paper with statements that Einstein could not fully comprehend. Einstein wrote Paul Hertz, "But I have shown in my paper that a usual gravitation law cannot be generally covariant. Do you not agree with this consideration?"[200] When writing this sentence Einstein apparently did not yet renounce the "Entwurf" field equations, however, he probably had serious doubts, because he was extremely nervous and cynical in his reply to Hertz. Subsequently he told Hertz that he does not remember he published that "dass die Welt auf sich selbst abwickelbar sein soll" (that the world will unwind upon itself). [201]

Einstein later explained in two letters the problems in his "Entwurf" theory up till then that led him to finally renounce the "Entwurf" equations – he sent these letters immediately after he presented to the Prussian Academy the final version of his general theory of relativity. The first letter on November 28, 1915 was sent to Arnold Sommerfeld. Einstein wrote,[202]

"I realized, namely, that my existing field equations of gravitation were entirely untenable! The following indications led to this:

1) I proved that the gravitational field on a uniformly rotating system does not satisfy the field equations.

2) The motion of Mercury's perihelion came to $18^0$ rather than $45^0$ per century.

3) The covariance considerations in my paper of last year do not supply the Hamiltonian function H. It leads, when it is properly generalized, to an arbitrary H. This showed that covariance with respect to 'adapted' coordinate system was a flop [ein Schlag ins Wasser war]".

The second letter was sent to Lorentz on January 1, 1916. Einstein wrote to Lorentz,[203]

"The gradual dawning realization of the incorrectness of the old gravitational field equations caused me last autumn hard times. I had already discovered earlier that Mercury's perihelion's motion was too small. In addition, I found that the equations were not covariant for substitutions which corresponded to a uniform rotation of the (new) reference system. Finally, I found that the consideration I made last year on the determination of Lagrange's H function for the gravitational field was absolutely illusory, in that it could easily be modified so that no limiting conditions had to be applied to H, so that it could have been chosen completely freely. Thus I came to the conviction that introducing the adapted coordinate system was a wrong path and that a more far-reaching covariance, preferably a *general* covariance, must be demanded".

**2. 3 September 30- October 1 1915: Metric field describing Rotating System is not a solution of "Entwurf" Field Equations, and Mercury Perihelion**

Einstein left for Zurich at the beginning of September for three weeks until the end of September. He stayed with his friend Heinrich Zangger.[204] Upon returning to Berlin he realized that the metric field describing a rotating system was not a solution of the "Entwurf" field equations. This realization could have stimulated what Einstein had "already discovered earlier that, Mercury's perihelion's motion was too small".[205]

Einstein's first known practical use of the "Entwurf" field equations was to see whether the "Entwurf" theory can account for the tiny discrepancy between the observed motion of Mercury and the motion predicted on the basis of Newton's theory of gravity; the discrepancy known as the anomalous advance of Mercury's perihelion. To solve this problem Einstein collaborated with his close friend Michele Besso in what is known by the name, the Einstein-Besso manuscript.[206]

In the Einstein-Besso manuscript of 1913 Einstein and Besso calculated the retrogression of the Perihelion of Mercury (pages 18 to 24) and retrogression of the nodes of the orbit of Mercury (pages 31 and 41-42) in the field of a rotating sun. [207] On pages 41-42 of the Einstein-Besso manuscript Einstein also checked whether the metric field describing a rotating system was a solution of the "Entwurf" field equations. Einstein could connect between the metric field describing the rotating sun and the metric in a rotating coordinate system.

Einstein checked whether he could use the "Entwurf" field equations to derive the 44-component of the Minkowski metric in rotating coordinates from its i4-components. Einstein wrote in a coordinate system rotating counterclockwise with respect to the z axis of an inertial coordinate system, a Minkowski metric field similar to the one below: [208]

| | | | |
|---|---|---|---|
| $-1$ | $0$ | $0$ | $-\omega y$ |
| $0$ | $-1$ | $0$ | $\omega x$ |
| $0$ | $0$ | $-1$ | $0$ |
| $-\omega y$ | $\omega x$ | $0$ | $1 - \omega^2(x^2 + y^2)$ |

$g_{14} = -\omega y$, $g_{24} = +\omega x$,

And the "Entwurf" field equation:

(I) $D_{\mu\nu}(g) + \kappa t_{\mu\nu} = 0$

$D_{44}(g) = -\Delta g_{44} - 4\omega^2$.

$$\kappa t_{\mu\nu} = \sum -\frac{1}{2}\frac{\partial g_{\tau\rho}}{\partial x_\mu}\frac{\partial \gamma_{\tau\rho}}{\partial x_\nu} + \frac{1}{4} g_{\mu\nu}\gamma_{\alpha\beta}\frac{\partial g_{\tau\rho}}{\partial x_\alpha}\frac{\partial \gamma_{\tau\rho}}{\partial x_\beta}$$

Einstein wrote a solution that has the form of the $g_{44}$ – the 44 component of Minkowski's metric in rotating coordinates, which was a solution of his equations, and made a sign error:

(II) $-\Delta g_{44} - 4\omega^2 + 2\omega^2 = 0$

$\Delta g_{44} = -2\omega^2$

Thus,

$-4\omega^2 + 2\omega^2 = -2\omega^2$.

Einstein thus thought during the period 1913-1915 that the metric field describing a rotating system was a solution of the "Entwurf" field equations, because of the little mistake he had made above on page 41.

The solution Einstein wrote for the $g_{44}$ was:

(1) $g_{44} = 1 - \omega^2(x^2 + y^2)$

Einstein asked himself whether the $g_{44}$ found above was the same as the $g_{44}$ one would obtain through direct transformation of the Minkowski metric to the rotating coordinate system. According to the above equation (1) Einstein arrived at the conclusion that it was, and next to the solution he wrote: "correct" ("stimmt").[209]

However, on page 42 Einstein redid the calculation. He put:

(III) $-\Delta g_{44} = (1/4)(4\omega^2)$.

[And thus, $-4\omega^2 + 2\omega^2 = (1/4)(4\omega^2)$].

The "Entwurf" equation is thus not satisfied and Einstein obtained a solution of the form for the $g_{44}$:

(2) $g_{44} = 1 - 3/4\ \omega^2(x^2 + y^2)$.

Einstein did not know what to do. He thus wrote the expression in the form:

$g_{44} = 1 - <3/4>\ \omega^2(x^2 + y^2)$.[210]

However, the above solution (1) obtained on page 41 from the "Entwurf" equations (while not noticing the sign error) perfectly fitted expectations of the principle of equivalence, and therefore Einstein finally chose (1) and not (2).

On September 30, 1915, Einstein was completely excited. He wrote Erwin Freundlich and thought the latter might help him, "I am writing you now about a scientific matter that electrifies me enormously. I have come upon a logical contradiction of a quantitative nature in the theory of gravitation, which proves to me that there must be a calculational error somewhere within my framework".[211]

Subsequently Einstein went straight to the discovery of the contradiction. Imagine an infinitely slowly rotating coordinate system, rotating with velocity ω. It can be shown by simple transformation that the gravitational field is given by the $g_{\mu\nu}$ metric field, the one Einstein had given two years earlier in the Einstein-Besso manuscript.

Einstein calculated from his "Entwurf" field equations the 44 component of the metric in rotating coordinates, equation (2). He then compared this to the 44 component $g_{44}$ obtained by direct transformation from the Galilean case, and he obtained equation (1). Einstein concluded that the difference was "*a blatant contradiction*" [ein flagranter Widerspruch]. [212]

Recall Einstein's introduction from his 1914 foundations "Entwurf" paper. Einstein said that the centrifugal force, which acts under given conditions upon a body, is determined by precisely the same natural constant as the effect of the gravitational field, such that we have no means to distinguish a "centrifugal field" from a gravitational field. He interpreted a rotating system as *at rest* and the centrifugal field as a gravitational field.[213] However, now it appeared as if this interpretation has crumbled down; it was like pulling a card (the rotation coordinate system solution) from the bottom of a stack and the entire structure of his field equations appeared to fall like a pile of cards. Einstein was in a crisis.

Moreover, as said above, for small angular velocities ($\omega \ll 1$) the metric above has the same general form as the metric field of the sun considered by Einstein and Besso in their manuscript in 1913, when calculating the perihelion of Mercury. Thus Einstein could use the approximation procedure used to find the metric field of the sun to second order to find $g_{44}$ for the Minkowski metric in a rotating coordinate system.[214]

He could also do the opposite; refer from his calculation of the rotating coordinate system about the case of the metric field of the sun and Mercury's perihelion.

Einstein could have done this in September 1915, because he told Freundlich that his solution for the Perihelion for Mercury (from 1913) was also problematic. He immediately related between the problem with rotation and the perihelion of Mercury:[215]

"I do not doubt, therefore, that the theory of perihelion motions is suffering from the same fault. Either the equations are already numerically incorrect (numerical coefficients), or I am applying the equations in principle in a wrong way".

Thereafter Einstein asked help from Freundlich: "I do not think that I myself am in the position to find the error, because my mind is locked in the same rut in this matter. Rather, I must depend on a person being with unspoiled brain matter to find the error. If you have time, do not forget to be occupied with the topic".[216] Freundlich came to visit Einstein and they might have dealt with Einstein's problem. Einstein began his letter to Otto Naumann from the first of October, 1915 by saying that Freundlich had visited him in the last days.[217]

Einstein redid the above calculation in the letter to Naumann: He first wrote equation (1) and the components $g_{14} = -\omega y$ and $g_{24} = \omega x$ of the Minkowski metric of the rotating coordinate system. Afterwards he wrote the "Entwurf" field equations (I), and equations (III). Equations (III) lead to the realization that the "Entwurf" equations are not satisfied:

(I) $D_{\mu\nu}(g) + \kappa t_{\mu\nu} = 0$

(III) $-\Delta g_{44} - 2\omega^2 - (1/4)(4\omega^2) \neq 0$ ,

$-\Delta g_{44} = 4\omega$.[218]

This was Einstein's "ein flagranter Widerspruch.[219]

Soon there was another nail put in the "Entwurf" coffin.

**2. 4 October 1915: Scalar H is invariant for linear transformations, covariance with respect to adapted coordinate systems was a flop**

In October1915 Einstein found a problem with the gravitational Lagrangian, the "Hamiltonian", H. Einstein wrote Lorentz on 12 October 1915 that he did not notice this mistake in 1914, because on page 1069 (section §13) of his 1914 paper he "carelessly introduced the condition that H was invariant with respect to linear transformations".

Einstein gave Lorentz the following explanation. He wrote (76) and (76a) as conditions for the adapted coordinate systems, and his "Entwurf" field equations (80a),[220] which are obtained by taking consideration of the gravitation tensor $\mathfrak{G}_{\mu\nu}$ and $Q = H\sqrt{-g}$. He then wrote the conservation equations (42c), Where $t_{\mu}^{\lambda}$ is given by (81b). Einstein demanded that the second term on the right-hand side of (80a) would be equal to the gravitational field energy tensor $t_{\mu}^{\lambda}$ multiplied by κ. By doing so Einstein arrived at the condition, equation (77) of his 1914 paper, $S_{\mu}^{\lambda} \equiv 0$. This is the condition for $H = \frac{Q}{\sqrt{-g}}$ (scalar) being an invariant with respect to *linear* substitutions. [from (77) by (76) one arrives at the conditions $B_{\mu} = 0$].

Einstein ended his letter to Lorentz by telling him that it follows from this invariance that H is a linear combination of five expressions, as he had found on page 1075 of his paper. With condition (77) Einstein equated H to the fourth one of these expressions (78). Einstein explained to Lorentz that he believed that the selection of H as the fourth expression (78) could be (formally) supported by the condition (77). However, he discovered that this was an error.[221] In fact the selection of H was not dependent on the condition (77); and H need not be limited. And thus covariance with respect to adapted coordinate systems was a flop.[222]

In October 1915 Einstein finally understood that his "Entwurf" gravitational field equations were problematic.

I am indebted to Prof John Stachel for his assistance and invaluable suggestions. It should be noted that the contents of this paper are the sole responsibility of the author.

## Endnotes

[1] Einstein, 1955, pp. 12-14.

[2] Einstein, 1912b, 1912c.

[3] Einstein, 1955, p. 15.

[4] Kollros, 1955, p. 27.

[5] Einstein, 1955, pp. 15-16.

[6] Einstein, 1955, p. 16.

---

[7] Ricci, Gregorio, and Levi-Civita, Tullio, "Méthodes de calcul differentiel absolu et leurs applications", *Mathematische Annalen* 54, 1901, pp. 125-201.

[8] Carl Friedrich Gauss's 1828 *Disquisitiones generales circa superficies curva.*

[9] Riemann, Bernhard, "On the hypotheses which lie at the Bases of Geometry". Translated by W. K. Clifford, *Nature* 8, 1873, pp. 183-184.

[10] Pais, 1982, p. 212.

[11] Kollros, 1955, p. 27; Christophel, E. B.,"Über die Transformation der homogenen Diffeentialausdrücke zweiten Grades", *Journal für die reine und angewandte Mathematik* 70, 1869, pp. 46-70.

[12] Einstein, 1955, pp. 10-11.

[13] Einstein, 1922, p. 47.

[14] Einstein to Sommerfeld, October 29, 1912, *CPAE*, Vol. 5, Doc. 421.

[15] Einstein, 1955, p. 15.

[16] *CPAE*, vol. 4, Doc. 10.

[17] *CPAE*, Vol.4, "Einstein on Gravitation and Relativity: The Collaboration with Marcel Grossmann", p. 296.

[18] Einstein and Grossmann, 1913. This paper is called by scholars the "Entwurf" paper. Actually Einstein himself also called this paper "Entwurf" (Einstein to Lorentz, 14 August, 1913, CPAE, Vol. 5, Doc. 467), and he and Grossmann referred to the theory presented in this paper as the "Entwurf" theory: in their second joint paper they wrote, "Gravitationsgleichung (21) bzw. (18) des 'Entwurfes'". Einstein and Grossmann, 1914, p. 217. The "Entwurf" theory was Einstein's theory of general relativity between spring 1913 and October 1915.

[19] Einstein and Grossman, 1914.

[20] *CPAE*, "Einstein on Gravitation and Relativity: The Collaboration with Marcel Grossman", Vol. 4, p. 294; Einstein and Grossmann, 1913 and 1914.

[21] *CPAE*, Vol.4, "Einstein on Gravitation and Relativity: The Collaboration with Marcel Grossmann", p. 294.

[22] Fritz Haber to Hugo Krüss, January 4, 1913, *CPAE*, Vol. 5, Doc. 428.

[23] *CPAE*, vol. 4, Doc. 10.

[24] Renn (2007), Vol. 2, pp. 451; 453.

[25] Janssen, Renn, Sauer, Norton and Stachel, 2007, in Renn (2007), pp. 681; 683; 704; Renn (2007), Vol. 2, pp. 462-465.

[26] Renn (2007), Vol. 2, pp. 466; 468.

[27] *CPAE*, Vol. 5. Note 1, p. 529.

[28] Seelig, 1954, pp. 172-173; Seelig, 1956, pp. 143-144.

[29] Proposal for Einstein's Membership in the Prussian Academy of Sciences, *CPAE*, Vol. 5, Doc. 445; Dokument Nr. 1 Wahlvorschlag fur A.Einstein zur Aufnahme als ordentliches Mitglied in die

Akademie d. Wiss. Von M. Planck, Berlin, 12 Juni 1913, in Kirsten and Treder, 1979, pp. 96-97. Other documents related to Einstein's position in Berlin are also found in this book.

[30] Kollros, 1955, in Seelig, 1956, pp. 29-30.

[31] *CPAE*, Vol. 5, note 4, p. 582.

[32] Haber to Krüss, January 4, 1913, *CPAE*, Vol. 5, Doc. 428, note. 5; Proposal for Einstein's Membership in the Prussian Academy of Sciences, June 12, 1913, *CPAE*, Vol. 5, Doc. 445, note 2.

[33] Dokument Nr. 2 Protokoll der Sitzung der Phys.-math. Klasse der Akademie d. Wiss. vom 12 Juni 1913, Auszung, in Kirsten and Treder, 1979, p. 97.

[34] CPAE, Vol. 5, note 4, p. 581.

[35] Einstein and Grossman, 1913.

[36] *CPAE*, Doc 14.

[37] Klein, 1970/1985, p. 294.

[38] *CPAE*, Vol. 5, note 4, p. 534.

[39] Frank, 1949/1979, p. 179; Frank, 1947/2002, p. 107; *CPAE*, Vol. 5, note 2, p. 534.

[40] Frank, 1949/1979, pp. 178-179; Frank, 1947/2002, pp. 107-108.

[41] Pais, 1982, p. 239.

[42] Frank, 1949/1979, pp. 179-180; Frank, 1947/2002, pp. 107-109.

[43] The Prussian Academy of Sciences to Einstein, November 22, 1913, *CPAE*, Vol. 5, Doc. 485.

[44] Kollros, 1955, in Seelig, 1956, pp. 29-30.

[45] Frank, 1949/1979, p. 181;Frank, 1947/2002, p. 108.

[46] Frank, 1949/1979, p. 182; Frank, 1947/2002, p. 108.

[47] Einstein to Elsa Löwenthal, December 2, 1913, *CPAE*, Vol. 5, Doc. 488.

[48] Einstein to Elsa Löwenthal, July 14?, 1913, *CPAE*, Vol. 5, Doc. 451.

[49] *CPAE*, Vol. 5, note 3, p. 534.

[50] *CPAE*, Vol. 5, note 3, p. 537.

[51] Einstein to Elsa Löwenthal, July 14?, 1913, *CPAE*, Vol. 5, Doc. 451.

[52] Einstein to Elsa Löwenthal, 14? July, 1913, *CPAE*, Vol. 5, Doc. 451, note 2.

[53] Einstein to Elsa Löwenthal, 19 July, 1913, *CPAE*, Vol. 5, Doc. 453.

[54] Einstein to Laub, 22 July, 1913, *CPAE*, Vol. 5, Doc. 455.

[55] Einstein to Elsa Löwenthal, after 19-before 24 July, 1913, *CPAE*, Vol. 5, Doc. 454.

[56] *CPAE*, Vol. 5, note 1, page 537.

[57] *CPAE*, Vol. 5, note 6, p. 538.

[58] Einstein to Lorentz, August 14, 1913, *CPAE*, Vol. 5, Doc. 467.

[59] Einstein to Elsa Löwenthal, August 11, 1913, *CPAE*, Vol. 5, Doc. 466.

[60] Einstein, 1913b; Einstein, 1913a.

[61] Einstein to Elsa Löwenthal, August 11?, 1913, *CPAE*, Vol. 5, Doc. 465.

[62] Einstein to Hopf, November 2, 1913, *CPAE*, Vol. 5, 1913, Doc. 480.

[63] Proposal for Einstein's Membership in the Prussian Academy of Sciences, *CPAE*, Vol. 5, Doc. 445.

[64] Dokument Nr. 6, Schreiben vom M. Planck an (?) im Kultasmin Berlin 7 Oktober 1913, in Kirsten and Treder, 1979, p. 100.

[65] Einstein to Robert Gnehm, November 30, 1913, *CPAE*, Vol. 5, Doc. 487.

[66] Note 1 to Doc. 494, *CPAE*, Vol. 5, p. 583.

[67] Einstein to the Prussian Academy of Sciences, December 7, 1913, *CPAE*, Vol. 5, Doc. 493.

[68] Einstein to the Prussian Academy of Sciences, December 7, 1913, *CPAE*, Vol. 5, Doc. 493.

[69] Robert Gnehm to Einstein, December 15, 1913, *CPAE*, Vol. 5, Doc. 494.

[70] *CPAE*, Vol. 5, note 4, p. 537.

[71] Kollros, 1955, in Seelig, 1956, pp. 29-30.

[72] Einstein to Elsa Löwenthal, February, 1914, CPAE, Vol. 5, Doc. 509. Note. 5.

[73] Einstein to Elsa Löwenthal, February, 1914, CPAE, Vol. 5, Doc. 509.

[74] Haber to Krüss, January 4, 1913, *CPAE*, Vol. 5, Doc. 428; *CPAE*, Vol. 5, note 9, p. 549.

[75] CPAE, Vol. 5, note. 5, p. 598.

[76] Einstein to Zangger, March 10, 1914, CPAE, Vol. 5, Doc. 513.

[77] Einstein to Lorentz, August 14, 1913, CPAE, Vol. 5, Doc 467.

[78] Einstein to Ehrenfest, second half of November, 1913, CPAE, Vol. 5, Doc. 484, note 3.

[79] Pais, 1982, p. 487.

[80] Einstein to Elsa Löwenthal, February, 1914, CPAE, Vol. 5, Doc. 509. Note. 5.

[81] Einstein to Elsa Löwenthal, after February, 1914, *CPAE*, Vol. 5, Doc. 509; Einstein to Elsa Löwenthal, after February 11, 1914, *CPAE*, Vol. 5, Doc. 510.

[82] Einstein to Zangger, March 10, 1914, *CPAE*, Vol. 5, Doc. 513; Einstein to Besso, March 10, 1914, *CPAE*, Vol. 5, Doc. 514.

[83] Einstein to Ehrenfest, before March 10, 1914, CPAE, Vol. 5, Doc. 512; note 2.

[84] Klein, 1970/1985, p. 296.

[85] Einstein to Elsa Löwenthal, after December 2, 1913, *CPAE*, Vol. 5, Doc. 489; note 3.

[86] *CPAE*, Vol. 5, Doc. 506, note. 9.

[87] Note 3, CPAE, Vol. 5, p. 604.

[88] Einstein, 1955, p. 16.

[89] Frank, 1949/1979, p. 182.

[90] Einstein to Besso, March 10, 1914, *CPAE*, Vol. 5, Doc. 514, note 9; Einstein to Freundlich, January 20, 1914, *CPAE*, Vol. 5, Doc. 506, note. 9.

[91] Einstein to Elsa Löwenthal, after December 2, 1913, *CPAE*, Vol. 5, Doc. 489.

[92] Einstein, 1934, p. 175; Einstein, 1954, p. 223.

It should be noted that Einstein's 1934 book *Mein Weltbild* was later edited and expanded in 1953 by Carl Seelig. And the 1954 English translation is from Einstein's *Mein Weltbild* edited by Seelig.

[93] Einstein to Planck, July 7, 1914, *CPAE*, Vol. 8, Doc. 18; Dokument Nr. 11, Brief von A. Einstein an M. Planck [Berlin] Dahlem, 7 Juli 1914, in Kirsten and Treder, 1979, pp. 14-105.

[94] Einstein to Zannger, March 10, 1914, *CPAE*, Vol. 5, Doc. 513.

[95] Einstein to Freundlich, January 20, 1914, *CPAE*, Vol. 5, Doc. 506.

[96] Einstein to Elsa Löwenthal, February, 1914, CPAE, Vol. 5, Doc. 509. Note. 5.

[97] Einstein to Freundlich, mid-August 1913, *CPAE*, Vol. 5, Doc. 468; Einstein to Freundlich, August 26, 1913, *CPAE*, Vol. 5, Doc. 472; George Hale to Einstein November 8, 1913, *CPAE*, Vol. 5, Doc. 483; Einstein to Freundlich, December 7, 1913, *CPAE*, Vol. 5, Doc. 492; Einstein to Freundlich, January 20, 1914, *CPAE*, Vol. 5, Doc. 506.

[98] Earman and Glymour, 1980, pp. 61-62.

[99] Earman and Glymour, 1980, p. 62.

[100] Einstein to Ehrenfest, August 19, 1914, *CPAE*, Vol. 8, Doc. 34.

[101] *CPAE*, Vol. 8, note 4. p. 57.

[102] Einstein, 1914b.

[103] Einstein and Grossmann, 1913, p. 7.

[104] Einstein to Sommerfeld, July 15, 1915, *CPAE*, Vol. 8, Doc. 96.

[105] Einstein, 1914b, p. 1031.

[106] Einstein, 1914b, p. 1032.

[107] Einstein, 1920b, p. 20, *CPAE*, Vol. 7, Doc. 31, p. 265

[108] .Einstein, 1920b, p. 21, *CPAE*, Vol. 7, Doc. 31, p. 265; Stachel, 2002/2007, pp. 262-263.

[109] Stachel, private conversation.

[110] .Einstein, 1920b, p. 21, *CPAE*, Vol. 7, Doc. 31, p. 265; Stachel, 2002/2007, pp. 262-263.

[111] Einstein, 1914b, pp. 1032-1033.

[112] Einstein, 1914b, pp. 1032-1033.

[113] Einstein, 1914b, p. 1033.

[114] Einstein, 1914b, p. 1034.

[115] Einstein, 1914b, p. 1037.

[116] Einstein, 1914b, p. 1035.

[117] Einstein, 1914b, p. 1035.

[118] Einstein, 1914b, p. 1044.

[119] Einstein, 1914b, p. 1046.

[120] Einstein, 1914b, p. 1046.

[121] Einstein, 1914b, p. 1046.

[122] Einstein, 1914b, p. 1046.

[123] Einstein, 1914b, p. 1052.

[124] Einstein, 1914b, p. 1054.

[125] Einstein, 1914b, p. 1055.

[126] Einstein, 1914b, p. 1055.

[127] Einstein, 1914b, pp. 1054-1055.

[128] Einstein, 1914b, p. 1056.

[129] Einstein, 1914b, p. 1056.

[130] Einstein, 1914b, pp. 1057-1058.

[131] Einstein, 1914b, p. 1058.

[132] Einstein, 1914b, p. 1059.

[133] Einstein, 1914b, p. 1066.

[134] Einstein, 1914b, p. 1067.

[135] Einstein, 1914b, p. 1067.

[136] Einstein, 1914b, pp. 1067-1068.

[137] Einstein, 1914b, p. 1069.

[138] Einstein, 1914b, pp. 1041-1042.

[139] Einstein, 1914b, p. 1069.

[140] Einstein and Grossmann, 1914, pp. 217; 224.

[141] Einstein and Grossmann, 1914, p. 219.

[142] Einstein, 1914b, p. 1070.

[143] Einstein, 1914b, p. 1070.

[144] Einstein, 1914b, p. 1071.

[145] Einstein, 1914b, p. 1071.

[146] Einstein, 1914b, p. 1071.

[147] Einstein, 1914b, pp. 1071-1072.

[148] Einstein, 1914b, p. 1073.

[149] Einstein, 1914b, p. 1071.

[150] Einstein, 1914b, p. 1073.

[151] Einstein, 1914b, p. 1073.

[152] Einstein, 1914b, pp. 1073-1074.

[153] Einstein, 1914b, p. 1074.

[154] Einstein, 1914b, pp. 1074-1076.

[155] Einstein, 1914b, p. 1076.

[156] Einstein, 1914b, pp. 1076-1077.

[157] Einstein, 1914b, p. 1077.

[158] Einstein, 1914b, pp. 1039-1040.

[159] Einstein, 1914b, pp. 1040-1041.

[160] Einstein, 1914b, pp. 1041-1042.

[161] Einstein, 1914b, p. 1042.

[162] Einstein, 1914b, pp. 1042-1043.

[163] Einstein, 1914b, p. 1043.

[164] Einstein, 1914b, p. 1043.

[165] Einstein, 1914b, pp. 1032-1033.

[166] Einstein, 1914b, p. 1044.

[167] Einstein, 1914b, p. 1058.

[168] Einstein, 1914b, p. 1050.

[169] Einstein, 1914b, p. 1051.

[170] Einstein, 1914b, p. 1049.

[171] Einstein, 1914b, p. 1051.

[172] Einstein, 1914b, p. 1052.

[173] Einstein, 1914b, p. 1052.

[174] Einstein, 1914b, p. 1053.

[175] Renn (2007), Vol. 2, pp. 451; 453.

[176] Einstein, 1914b, p. 1053.

[177] Einstein, 1914b, p. 1053.

[178] Einstein, 1914b, p. 1053.

[179] Einstein, 1914b, p. 1054.

[180] Einstein, 1914b, p. 1056.

[181] Einstein, 1914b, p. 1055.

[182] Einstein, 1914b, p. 1055.

[183] Cattani and De Maria, 1989, in Howard and Stachel (1989), p. 176.

[184] Einstein to Levi-Civita, April 2, 1915, *CPAE*, Vol. 8, Doc. 69.

[185] Einstein to Levi-Civita, March 5, 1915, *CPAE*, Vol. 8, Doc. 60; Einstein to Levi-Civita, March 17, 1915, *CPAE*, Vol. 8, Doc. 62; Einstein to Levi-Civita, March 20, 1915, *CPAE*, Vol. 8, Doc. 64; Einstein to Levi-Civita, March 26, 1915, *CPAE*, Vol. 8, Doc. 66; Levi-Civita to Einstein, March 28, 1915, *CPAE*, Vol. 8, Doc. 67; Einstein to Levi-Civita, April 2, 1915, *CPAE*, Vol. 8, Doc. 69; Einstein to Levi-Civita, April 8, 1915, *CPAE*, Vol. 8, Doc. 71; Einstein to Levi-Civita, April 11, 1915, *CPAE*, Vol. 8, Doc. 74; Einstein to Levi-Civita, April 14, 1915, *CPAE*, Vol. 8, Doc. 75; Einstein to Levi-Civita, April 20, 1915, *CPAE*, Vol. 8, Doc. 77; Einstein to Levi-Civita, April 21, 1915, *CPAE*, Vol. 8, Doc. 78; Einstein to Levi-Civita, May 5, 1915, *CPAE*, Vol. 8, Doc. 80.

[186] *CPAE*, Vol. 8, note 5, p. 146.

[187] Appendix B, "*Nachschrift* of Einstein's Wolfskehl lectures, Summer 1915", *CPAE*, Vol. 6, pp. 586-590.

[188] Appendix B, "*Nachschrift* of Einstein's Wolfskehl lectures, Summer 1915", *CPAE*, Vol. 6, p. 589.

[189] Appendix B, "*Nachschrift* of Einstein's Wolfskehl lectures, Summer 1915", *CPAE*, Vol. 6, p. 589.

[190] Einstein to Zangger, July 7, 1915, *CPAE*, Vol. 8, Doc. 94.

[191] Levi-Civita, T., and Ricci-Curbastro, G., "Méthodes de calcul différential absolu et leurs applications", *Mathematische Annalen* 54 (1900), pp. 125–201.

[192] Einstein, 1911a.

[193] Einstein, 1914b.

[194] Einstein to Sommerfeld, July 15, 1915, *CPAE*, Vol. 8, Doc. 96.

[195] Einstein to Zangger, Between July 24 and August 7, 1915, *CPAE*, Vol. 8, Doc. 101.

[196] Wander de Hass was a Dutch physicist (Lorentz's son-in-law, Geertruida was Lorentz's elder daughter) who worked with Einstein on the "Einstein-De Hass Effect".

Einstein, Albert, and de Haas, Wander, J., "Experimenteller Nachweis der Ampereschen Molekularstörme", *Deutsche Physikalische Gesellschaft, Verhandlungen* 17, 1915, pp. 152–170.

[197] Einstien to Wander and Geertruida de Hass, August 16, 1915, *CPAE*, Vol. 8, Doc. 110.

[198] Einstein to Hilbert, November 7, 1915, *CPAE*, Vol. 8, Doc. 136.

[199] Einstein to Hilbert, November 7, 1915, *CPAE*, Vol. 8, Doc. 136.

[200] Einstein to Hertz, between August 14 and October 9, 1915, *CPAE*, Vol. 8, Doc. 108.

[201] Einstein to Hertz, between August 14 and October 9, 1915, *CPAE*, Vol. 8, Doc. 108.

[202] Einstein to Sommerfeld, November 28, 1915, *CPAE*, Vol. 8, Doc. 153.

[203] Einstein to Lorentz, January, 1 1916, *CPAE*, Vol. 8, Doc. 177.

[204] Einstein to Hertz, August 22, 1915, *CPAE*, Vol. 8, Doc. 111; Fölsing, 1997, p. 365.

[205] Einstein to Lorentz, January, 1 1916, *CPAE*, Vol. 8, Doc. 177.

[206] *CPAE*, Doc 14.

[207] *CPAE*, Vol. 4, "The Einstein-Besso Manuscript on the Motion of the Perihelion of Mercury", pp. 347-348.

[208] *CPAE*, Vol. 4, Doc. 14, p. 41; *CPAE*, Vol. 4, Doc. 14, note 186, p. 445.

[209] *CPAE*, Vol. 4, Doc. 14, p. 41; *CPAE*, Vol. 4, Doc. 14, note 186, p. 445; *CPAE*, Vol. 4, Doc. 14, note 184, p. 443.

[210] *CPAE*, Vol. 4, Doc. 14, p. 42; *CPAE*, Vol. 4, Doc. 14, note 193, p. 447.

[211] Einstein to Freundlich, September 30, 1915, *CPAE*, Vol. 8, Doc. 123.

[212] Einstein to Freundlich, September 30, 1915, *CPAE*, Vol. 8, Doc. 123.

[213] Einstein, 1914b, p. 1032.

[214] *CPAE*, Vol. 4, Doc. 14, p. 41; *CPAE*, Vol. 4, Doc. 14, note 185, p. 443.

[215] Einstein to Freundlich, September 30, 1915, *CPAE*, Vol. 8, Doc. 123.

[216] Einstein to Freundlich, September 30, 1915, *CPAE*, Vol. 8, Doc. 123.

[217] Einstein to Naumann, October 1, 1915, *CPAE*, Vol. 8, Doc. 124; *CPAE*, Vol. 8, note 3, p. 178.

[218] *CPAE*, Vol. 8, note 5, p. 180.

[219] Einstein to Freundlich, September 30, 1915, *CPAE*, Vol. 8, Doc. 123.

[220] Einstein to Lorentz, October 12, 1915, *CPAE*, Vol. 8, Doc. 129.

[221] Einstein to Lorentz, October 12, 1915, *CPAE*, Vol. 8, Doc. 129.

[222] Einstein to Sommerfeld, November 28, 1915, *CPAE*, Vol. 8, Doc. 153.